\documentclass{elsarticle}
\usepackage{amsmath}
\usepackage{amssymb}
\usepackage{cases}
\usepackage{enumerate}
\usepackage{color}
\usepackage{graphicx}
\usepackage{lineno,hyperref}

\journal{}

\newcommand{\vep}{\varepsilon}

\newcommand{\be}{\begin{equation}}
\newcommand{\ee}{\end{equation}}
\newcommand{\ba}{\begin{array}}
\newcommand{\ea}{\end{array}}
\newcommand{\bea}{\begin{eqnarray}}
\newcommand{\eea}{\end{eqnarray}}
\newcommand{\beas}{\begin{eqnarray*}}
\newcommand{\eeas}{\end{eqnarray*}}

\newtheorem{remark}{Remark}[section]
\newtheorem{theorem}{Theorem}[section]
\newtheorem{lemma}{Lemma}[section]

 \newcommand{\bx}{{\bf x} }
\DeclareMathOperator*{\essinf}{ess\,inf}

\renewcommand{\ldots}{\dotsc}

\newcommand{\tphi}{{\tilde{\phi}} }
\newcommand{\trho}{{\tilde{\rho}} }

\begin{document}

\begin{frontmatter}

\title{Ground states  of Bose-Einstein condensates with higher order interaction}

\author{Weizhu Bao}
\address{Department of Mathematics,  National University of Singapore, Singapore 119076, Singapore}
\ead{matbaowz@nus.edu.sg}
\fntext[myfootnote]{Url:{\it{http://www.math.nus.edu.sg/\~{}bao/}}}

\author{Yongyong Cai\corref{cor1}}
\cortext[cor1]{Corresponding author}
\address{Beijing Computational Science Research Center,  No. 10 West Dongbeiwang Road, Haidian District, Beijing 100193,
P. R. China}
\ead{yongyong.cai@csrc.ac.cn}

\author{Xinran Ruan\corref{}}
\address{Department of Mathematics,  National University of Singapore, Singapore 119076, Singapore}
\ead{a0103426@u.nus.edu}

\begin{abstract}
We analyze the ground state of a Bose-Einstein condensate in the presence of higher-order interaction (HOI),
modeled by a modified Gross-Pitaevskii equation (MGPE). In fact, due to the appearance of HOI,
the ground state structures become very rich and complicated.
We establish the existence and  non-existence
results under different parameter regimes, and obtain their limiting behaviors and/or structures
with different combinations of HOI and contact interactions. Both the whole space case and  the bounded domain case are considered, where
different structures of ground states are identified.
\end{abstract}

\end{frontmatter}


\section{Introduction}
\label{s1} \setcounter{equation}{0}
Bose-Einstein condensates (BECs) have been thoroughly studied since its first experimental realization in 1995 \cite{Anderson,Davis} and many of its properties have been investigated theoretically  based on the mean-field  Gross-Pitaevskii equation (GPE).
In the derivation of GPE, one key assumption is
 that the binary interaction between the particles can be
well described by  the shape-independent approximation (or pseudopotential approximation), i.e. a Dirac function, where the interaction strength
is characterized by  the $s$-wave scattering length \cite{Esry}. It is well-known  that such approximation is valid in
low energies (or low densities) and becomes less valid in  high energies (or high densities). Therefore, numerous efforts have been devoted
to the improvements of the pseudopotential approximation for the two-body interaction, which lead to better mean field theory towards the understanding of
BEC experiments.

In \cite{Esry,Collin}, a higher order interaction (HOI) correction to the pseudopential approximation has been analyzed. As a consequence,
at temperature $T$ much smaller than the critical temperature $T_c$, a BEC with HOI   can be described by the
wave function $\psi:=\psi(\bx,t)$
whose evolution is governed by the  dimensionless modified Gross-Pitaevskii equation (MGPE) in  three dimensions (3D) \cite{Ruan,Collin,Esry}
\be\label{eq:mgpe3d}
i\partial_t \psi=\left[-\frac12\nabla^2
+V(\bx)
+ g_{0}|\psi|^2-g_1\nabla^2|\psi|^2\right]\psi,
 \ee
where $t$ is time,
$\bx=(x,y,z)^T\in \mathbb R^3$   is the Cartesian coordinate,
$g_{0}$ is the contact interaction constant (positive for
repulsive interaction and negative for attractive interaction), $g_{1}$ is the constant describing the higher order
correction of the contact interaction due to the finite size effects,
and $V(\bx)$ is a given
real-valued external trapping potential
and is commonly chosen to be the harmonic potential in typical experiments as
\begin{equation}\label{eq:pot}
V(\bx)=\frac{1}{2}\left(\gamma_x^2x^2+\gamma_y^2y^2+\gamma_z^2z^2\right),\quad \bx\in{\mathbb R}^3.
\end{equation}



When the trapping potential in (\ref{eq:pot}) is strongly anisotropic, i.e. $\gamma_x,\gamma_y\ll\gamma_z$ for a quasi-2D BEC or $\gamma_x\ll\gamma_y,\gamma_x\ll\gamma_z$ for a quasi-1D
BEC, similar to
the dimension reduction of the conventional GPE for a BEC \cite{Bao2013,Bao2014,Ben,PitaevskiiStringari},
the MGPE (\ref{eq:mgpe3d}) in 3D can be formally reduced
to two dimensions (2D) or one dimension (1D) for the  disk-shaped or cigar-shaped BEC \cite{Ruan, Veksler}, respectively.
In fact, the resulting MGPE can be written
in a unified form in $d$-dimensions ($d=1,2,3$) with $\bx\in \mathbb R^d$  and $\bx=x\in\mathbb R$ for $d=1$,
$\bx=(x,y)^T\in\mathbb R^2$ for $d=2$ and
$\bx=(x,y,z)^T\in\mathbb R^3$   for $d=3$ as
\be\label{eq:mgpe}
 i\partial_t \psi=\left[-\frac{1}{2}\nabla^2
+V(\bx)
+\beta|\psi|^2-\delta\nabla^2|\psi|^2\right]\psi,
 \ee
where
\be\label{eq:potential:ho}
V(\bx)=\begin{cases}
\frac12(\gamma_x^2x^2+\gamma_y^2y^2+\gamma_z^2z^2),& d=3,\\
\frac12(\gamma_x^2x^2+\gamma_y^2y^2),& d=2,\\
\frac12\gamma_x^2x^2, &d=1.
\end{cases}
\ee

For other potentials such as box potential, optical lattice potential and double-well potential,
we refer to \cite{Bao2013,PitaevskiiStringari} and references therein.
Thus, in the subsequent discussion, we will treat
the external potential $V(\bx)$ in (\ref{eq:mgpe})
as  a general real-valued function and the parameters $\beta$ and $\delta$  as
arbitrary real constants.  In addition,
without loss of generality, we assume $V(\bx)\ge0$  
in the rest of this paper.
The dimensionless MGPE (\ref{eq:mgpe}) conserves the total mass
 \be\label{eq:mass}
N(t):=\|\psi(\cdot,t)\|^2=\int_{{\mathbb R}^d}|\psi(\bx,t)|^2d\bx
\equiv \|\psi(\cdot,0)\|^2=1, \quad t\ge0, \ee
and the energy per particle
\be \label{eq:energy}
E(\psi)=\int_{{\mathbb R}^d}\biggl[
\frac12|\nabla\psi|^2+V(\bx)|\psi|^2+\frac{\beta}{2}|\psi|^4+\frac{\delta}{2}\left|\nabla|\psi|^2\right|^2\biggl]\,d\bx.
\ee
Theoretically, other higher order terms  can be included in the MGPE \eqref{eq:mgpe} as the higher order corrections of
the two-body interaction \cite{Collin}. Here, we focus on the current MGPE \eqref{eq:mgpe} to understand the idea behind the theory.
In fact, MGPE \eqref{eq:mgpe} has  been found in other applications (in a generalized form), such as the modeling of
ultrashort laser pulses in plasmas \cite{Borov,De}, description of the thin-film superfluid condensates  \cite{Kuri}, study of the
Heisenberg ferromagnets \cite{Takeno}.
MGPE \eqref{eq:mgpe} with $\delta=0$ has been thoroughly studied in the literature
and we refer the readers to \cite{Bao2013,Bao2014,PitaevskiiStringari} and reference therein.
However, there have been only a few mathematical results for
 MGPE \eqref{eq:mgpe}, including the local well-posedness of the Cauchy problem \cite{Popp,Marz}, existence
 of  solutions to the time independent version of \eqref{eq:mgpe} \cite{Liu,Liu2}, the stability of standing waves \cite{ColinM},
 spectral method for \eqref{eq:mgpe} \cite{Lu}, etc.
 To the best of our knowledge, all the known mathematical results for MGPE \eqref{eq:mgpe}
 are not based on the BEC  applications, and only some physical studies are available for MGPE \eqref{eq:mgpe} originating from BEC,
 like the ground state properties \cite{Fu,Tho} and the dynamical instabilities \cite{Qi,Qix}. In \cite{Ruan}, we have studied the dimension reduction of MGPE in lower dimensions.
 Here, we will present
our mathematical results on ground states of BEC based on
the MGPE \eqref{eq:mgpe}.
In particular, much effort will be devoted to the study of the existence and qualitative properties as well as the asymptotic profiles of the ground state under different parameter regimes.

 The paper is organized as follows. In section 2, we
establish existence, uniqueness and non-existence
results of ground states under different parameter regimes as well as qualitative properties including regularity and decay of the ground state in the far field.
We study the asymptotic profiles of ground states in different parameter regimes under a harmonic potential in section 3 and under a box  potential in section 4. In particular, we are interested in the regimes with vanishing $\delta$ ($\delta\to0^+$) and large interactions $\delta\to+\infty$, $|\beta|\to\infty$.
 Some conclusions are drawn in section 5.

\section{Mathematical analysis of the ground state}
In this section, we focus on the existence and uniqueness of the ground states as well as the qualitative properties such as regularity and far field decay.
\subsection{Existence and uniqueness}
\label{s2} \setcounter{equation}{0}
Introduce the function space
\begin{equation*}
X=\left\{\phi\in H^1(\mathbb R^d)\left| \|\phi\|_X^2=
\|\phi\|^2+\|\nabla\phi\|^2+\int_{\mathbb R^d}V(\bx)|\phi(\bx)|^2\,d\bx<\infty\right.\right\}.
\end{equation*}
The ground state $\phi_g:=\phi_g(\bx)$ of
a BEC modeled by the MGPE (\ref{eq:mgpe})
is defined as the minimizer of the energy functional (\ref{eq:energy}) under the
constraint (\ref{eq:mass}), i.e.

  Find $\phi_g \in S$ such that
\begin{equation}\label{eq:minimize}
    E_g := E\left(\phi_g\right) = \min_{\phi \in S}
    E\left(\phi\right),
  \end{equation}
where $S$ is  defined as \be\label{eq:nonconset}
S:=\left\{\phi \in X| \,\|\phi\|=1,\quad E(\phi)<\infty\right\}.\ee
In addition, the ground state $\phi_g$ is a solution to the following nonlinear
eigenvalue problem, i.e. the Euler-Lagrange equation of the problem (\ref{eq:minimize})
\be\label{eq:e-l}
\mu \phi=\left[-\frac{1}{2}\nabla^2
+V(\bx)
+\beta|\phi|^2-\delta\nabla^2|\phi|^2\right]\phi,
 \ee
under the normalization constraint $\phi\in S$, where the corresponding eigenvalue (or chemical potential)  $\mu:=\mu(\phi)$ can be computed as
 (multiply \eqref{eq:e-l} by $\phi$ and integrate over $\bx$)
\be
\mu=E(\phi)+\int_{\mathbb R^d}\left(\frac{\beta}{2}|\phi|^4+\frac{\delta}{2}\left|\nabla|\phi|^2\right|^2
\right)\,d\bx.
\ee

The following embedding results hold \cite{Bao2013}.
\begin{lemma}\label{lem:compact} Under the assumption that $V(\bx)\ge0$
for $\bx\in {\mathbb R}^d$ is a confining potential,
i.e. $\lim\limits_{R\to\infty}\essinf_{|\bx|<R} V(\bx)=\infty$,
we have that
the embedding $X\hookrightarrow L^{p}(\mathbb R^d)$
is compact provided that exponent $p$
 satisfies
\begin{equation}
\begin{cases}
p\in[2,6),\quad d=3,\\
p\in[2,\infty),\quad d=2,\\
p\in [2,\infty],\quad d=1.
\end{cases}
\end{equation}
\end{lemma}

The existence and uniqueness of the ground state when $\delta=0$ has been thoroughly studied in \cite{Bao2013,Lie,Guo,Guo1}.
When $\delta\neq0$, we have
\begin{theorem}\label{thm:mres}(Existence and uniqueness)
Suppose  $\delta\neq0$ and $V(\bx)\ge 0$ satisfies the confining condition, i.e.
$\lim\limits_{|\bx|\to\infty}V(\bx)=\infty$, then  there exists a
minimizer
 $\phi_g\in S$ of
 \eqref{eq:minimize} if and only if $\delta>0$. 
Furthermore, $e^{i\theta}\phi_g$ is also a ground state
of \eqref{eq:minimize} for any $\theta\in\mathbb [0,2\pi)$.
The ground state  $\phi_g$ can be chosen as non-negative $|\phi_g|$ and
the non-negative ground state is unique
if $\delta\ge0$ and $\beta\ge0$.

 The uniqueness result can be generalized to the case with negative $\beta$ when the problem is defined on a bounded connected open domain $\Omega$, i.e. the potential $V(\bx)=+\infty$ for $\bx\notin\Omega$.
In such case, the zero Dirichlet boundary conditions on $\partial\Omega$ are imposed for the wave functions, and
for any $\delta>0$, there exists $C_{\Omega}>0$ (depending on $\Omega$) such that when $\beta> -\delta/C_{\Omega}$, the non-negative ground state $\phi_g\in H_0^1(\Omega)$ of \eqref{eq:minimize}
is unique.
\end{theorem}

{\noindent\it Proof}.

(i)
We start with the  existence. 
Assume $\delta>0$,
 by the inequality \cite{LiebL}
\be
|\nabla|\phi(\bx)||\leq |\nabla\phi(\bx)|,\quad \text{a.e.}\quad \bx\in\Bbb R^d,
\ee
we deduce
\be
E(\phi)\ge E(|\phi|),
\ee
where equality holds iff $\phi=e^{i\theta}|\phi|$ for some constant $\theta\in[0,2\pi)$. Therefore, it suffices to consider the real non-negative minimizers of
\eqref{eq:minimize}.  On the other hand, Nash inequality $\|f\|_{L^2}^{1+2/d}\leq C\|f\|_{L^1}^{2/d}\|\nabla f\|_{L^2}$ and Young's inequality imply that for $\rho=|\phi|^2$ ($\phi\in S$),
\begin{align*}
\int_{\Bbb R^d}|\phi|^4\,d\bx\leq (C\int_{\Bbb R^d}\rho(\bx)\,d\bx)^{4/d+2}\|\nabla|\phi|^2\|^{2d/d+2}
\leq  \frac{C}{\vep}+\vep\|\nabla\rho\|^2,\quad \forall \vep>0,
\end{align*}
and we can conclude that $E(\phi)$ ($\phi\in S$) is bounded from below
\begin{equation*}
E(\phi)\ge \int_{\Bbb R^d}\left(\frac12|\nabla\phi|^2+V(\bx)|\phi|^2+\frac{\delta}{4}\left|\nabla|\phi|^2\right|^2\right)\,d\bx-C.
\end{equation*}
Taking a nonnegative minimizing sequence $\{\phi_n\}_{n=1}^\infty\subset S$, we find the $\phi_n$ is uniformly bounded in $X$ and  there exists $\phi_\infty \in X$ and a subsequence
(denote as the original sequence for simplicity) such that
\be
\phi_n\rightharpoonup
 \phi_\infty \quad \text{in}\quad X,
\ee
Lemma \ref{lem:compact} ensures that $\phi_n \to \phi_\infty$ in $L^p$ with $p$ given in the lemma and so $\nabla |\phi_n|^2$ converges to $\nabla|\phi_\infty|^2$ in the sense of distribution.  Noticing that $\|\nabla|\phi_n|^2\|$ is uniformly bounded and hence $\nabla|\phi_n|^2$ converges weakly in $L^2$ and we then get
$\nabla |\phi_n|^2\rightharpoonup
\nabla |\phi_\infty|^2\quad \text{in}\quad L^2$.
 Thus we know $\phi_\infty\in S$ with $\phi_\infty$ being nonnegative, and under the condition $\delta>0$,
\be
E(\phi_\infty)\leq \liminf_{n\to\infty} E(\phi_n)=\min_{\phi\in S} E(\phi),
\ee
which shows that $\phi_\infty$ is a ground state.

Secondly, for the case $\beta>0$ and $\delta>0$, we prove the uniqueness of the nonnegative ground state. Denote $\rho=|\phi|^2$, then for $\phi=\sqrt{\rho}\in S$, the energy is
\be
E(\sqrt{\rho})=\int_{{\mathbb R}^d}\biggl[
\frac12|\nabla\sqrt{\rho}|^2+V(\bx)\rho+\frac{\beta}{2}|\rho|^2+\frac{\delta}{2}\left|\nabla\rho\right|^2\biggl]\,d\bx.
\ee
The sum of first three terms of the energy  is strictly convex in $\rho$ \cite{Lie,Bao2013}, and the last term is also convex because it is  quadratic in $\rho$ and $\delta>0$. Hence,
we know $E(\sqrt{\rho})$ is strictly convex in $\rho$ and the uniqueness of the nonnegative ground state follows \cite{Lie,Bao2013}. 

When $\delta<0$, we show the nonexistence of the ground state.
Choosing a non-negative smooth function $\varphi(\bx) \in S$ with compact support  and denoting $\varphi_\vep(\bx)=\vep^{-d/2}\varphi(\bx/\vep) \in S$, we have
\be
E(\varphi_\vep)=\int_{\Bbb R^d}\left[\frac{1}{2\vep^2}|\nabla\varphi|^2+ V(\vep\bx)|\varphi|^2+\frac{\beta}{2\vep^d}|\varphi|^4+\frac{\delta}{2\vep^{2+d}}\left|\nabla|\varphi|^2\right|^2\right]\,d\bx.
\ee
From the above equation, we see that $\lim\limits_{\vep\to 0^+}E(\varphi_\vep)\to-\infty$ if $\delta<0$ and there exists no ground state.

(ii) For problems defined on a bounded connected open domain,
we have $\phi_g\in H_0^1(\Omega)$. Using Poincar\'{e} inequality, 
there exists $C_{\Omega}>0$ such that
\be\label{eq:sobo1}
\|f\|_{L^2(\Omega)}\leq C_{\Omega}\|\nabla f\|_{L^2(\Omega)}.
\ee
Denote $\rho=|\phi|^2$, then for $\phi=\sqrt{\rho}\in S$,  and we claim the energy $E(\sqrt{\rho})$ is convex in $\rho$ for $\beta\ge-\delta/C_{\Omega}$. To see this,
we only need examine the case $\beta\in (-\delta/C_{\Omega},0)$. For any $\sqrt{\rho_j}\in S$ with $\rho_j\in H_0^1(\Omega)$ and $\theta\in[0,1]$, we have
\begin{align*}
&\theta E(\sqrt{\rho_1})+(1-\theta)E(\sqrt{\rho_2})-E(\sqrt{\theta\rho_1+(1-\theta)\rho_2})
\\ \ge&\frac12\theta(1-\theta)\left(\beta\|\rho_1-\rho_2\|^2+\delta\|\nabla(\rho_1-\rho_2)\|^2\right)\\
\ge&\frac12\theta(1-\theta)\left(-\delta\|\nabla(\rho_1-\rho_2)\|^2+\delta\|\nabla(\rho_1-\rho_2)\|^2\right)=0,
\end{align*}
where we used the fact $\|\nabla\sqrt{\rho}\|^2$ is convex in $\rho$. This shows $E(\sqrt{\rho})$ is convex   when $
\beta>-\frac{\delta}{C_{\Omega}}$. The uniqueness follows. $\hfill\Box$
\begin{remark}
In the general whole space case,  the energy functional $E(\sqrt{\rho})$ is no longer convex and the uniqueness when $\beta<0$
is not clear (see recent results obtained by Guo et al. in \cite{Guo} about the uniqueness when $\delta=0$ with small $|\beta|$).
\end{remark}

\subsection{Regularity and decay}
Concerning the ground state of \eqref{eq:minimize}, we have the following properties.
\begin{theorem} \label{thm:gsprop} Let $\delta>0$ and $\phi_g\in S$ be the nonnegative ground state of \eqref{eq:minimize}, we have the following properties:

(i) There exists $\alpha>0$ and $C>0$ such that $|\phi_g(\bx)|\leq C e^{-\alpha |\bx|}$, $\bx\in\Bbb R^d$.

(ii) If $V(\bx)\in L^\infty_{\rm loc}(
\Bbb R^d)$, we have  $\phi_g$ is once continuously differentiable and
 $\nabla \phi_g$ is H\"older continuous with order $1$. In particular, if $V(\bx)\in C^\infty$, $\phi_g$ is smooth.
\end{theorem}
{\noindent\it Proof}. (i) We show the $L^\infty$ bound of $\phi_g$ by a Moser's iteration and De Giorgi's iteration following \cite{Liu2}. From the fact that $\phi_g\in S$
minimizes energy \eqref{eq:energy}, it is easy to check that $\phi_g$ satisfies the Euler-Lagrange equation \eqref{eq:e-l}, which shows that for any test function $\varphi\in C_0^\infty(\Bbb R^d)$, the following holds for $\phi=\phi_g$
\begin{equation}\label{eq:weak}
\int_{\Bbb R^d}\left[\frac{1}{2}\nabla\phi\cdot\nabla\varphi
+V(\bx)\phi\varphi
+\delta\phi\nabla \phi\cdot\nabla (\phi\varphi)\right]\,d\bx=\int_{\Bbb R^d}\left[-\beta|\phi|^2\phi\varphi+\mu\phi\varphi\right]\,d\bx.
\end{equation}

Using the Moser and De Giorgi iterations, we will prove that any
 weak solution $\phi\in X\cap\{E(\phi)<\infty\}$ of \eqref{eq:weak}
is bounded and decays exponentially as $|\bx|\to\infty$.  In detail, we first observe that by an approximation argument,
the test function $\varphi$ can be any functions in $X$ such that
$\int_{\Bbb R^d}|\varphi|^2|\nabla\phi|^2\,d\bx<\infty$ and $\int_{\Bbb R^d}|\phi|^2|\nabla\varphi|^2\,d\bx<\infty$.

Firstly, we show that   for all $q\ge1$, $\int_{\Bbb R^d}(1+\phi^{2q})|\nabla\phi|^2\,d\bx<\infty$.
 Choose $q_0=12$, since $\nabla\phi^2\in L^2$ and $\phi\in H^1$, $\phi\in L^{p}(\Bbb R^d)$ ($\forall p\in[2,q_0],\quad d=1,2,3$).
 Let $\varphi=|\phi_M|^{q_0-4}\phi_M$ ($M>0$) be the test function, where $\phi_M(\bx)=\phi(\bx)$ if $\bx\in\{|\phi(\bx)|\leq M\}$,
and $\phi_M(\bx)=\pm M$ if $\bx\in\{\phi(\bx)\gtrless \pm M\}$. Plugging $\varphi=|\phi_M|^{q_0-4}\phi_M$ into \eqref{eq:weak}, we obtain
\begin{align*}
&(q_0-3)\int_{\Bbb R^d}(\frac12+\delta\phi^2)|\phi_M|^{q_0-4}\nabla\phi\cdot\nabla\phi_M\,d\bx+\delta\int_{\Bbb R^d}\phi\phi_M|\phi_M|^{q_0-4}|\nabla\phi|^2\,d\bx\\
&+\int_{\Bbb R^d}V(\bx)\phi\phi_M|\phi_M|^{q_0-4}\,d\bx=
\int_{\Bbb R^d}\left(-\beta|\phi|^2\phi+\mu\phi\right)|\phi_M|^{q_0-4}\phi_M\,d\bx.
\end{align*}
Letting $M\to\infty$, we get
\begin{equation}\label{eq:test}
(q_0-2)\delta\int_{\Bbb R^d}|\phi|^{2\tilde{q}}|\nabla\phi|^2\,d\bx+\int_{\Bbb R^d}V(\bx)|\phi|^{2\tilde{q}}\,d\bx\leq
\int_{\Bbb R^d}\left(|\beta| |\phi|^{q_0}+|\mu||\phi|^{q_0-2}\right)\,d\bx,
\end{equation}
which shows $\int_{\Bbb R^d}|\phi|^{{2\tilde{q}}}|\nabla\phi|^2\,d\bx<\infty$ ($\tilde{q}=\frac{q_0}{2}-1$). So $\nabla \phi^{\tilde{q}+1}\in L^2$
 and for $q_1=6(\tilde{q}+1)=3q_0=36$, $\phi\in L^{p}(\Bbb R^d)$ ($\forall p\in[2,q_1],\quad d=1,2,3$). Then, the Moser
iteration can continue with $q_j=3^jq_0$, and $\phi\in L^{q_j}(\Bbb R^d)$ (it is obvious when $d=1,2$) which verifies our claim.
In particular $\phi\in L^p$ for any $p\in[2,\infty)$.

Secondly, we show that $\phi\in L^\infty(\Bbb R^d)$ and $\lim_{|\bx|\to\infty}\phi(\bx)=0$ by De Giorgi's iteration.
Denoting $f=-\beta|\phi|^2\phi+\mu\phi$ and choosing
test function $\varphi(\bx)=(\xi(\bx))^2(\phi(\bx)-k)_+$ with $k\ge0$ in \eqref{eq:weak}, where $(\phi(\bx))_+=\max\{\phi,0\}$  and $\xi(\bx)$ is
a smooth cutoff function,  we have
\begin{align*}
&\int_{\Bbb R^d} \left[(\frac12+\delta\phi^2+\delta\phi(\phi-k)_+)|\xi|^2|\nabla(\phi-k)_+|^2+ V(\bx)|\xi|^2\phi(\phi-k)_+\right]\,d\bx\\
&=\int_{\Bbb R^d}\left[-(1+2\delta\phi^2)(\phi-k)_+\xi\nabla(\phi-k)_+\cdot\nabla\xi+f\xi^2(\phi-k)_+\right]\,d\bx.
\end{align*}
Cauchy inequality gives that
\begin{align*}
&\int_{\Bbb R^d}-(1+2\delta\phi^2)(\phi-k)_+\xi\nabla(\phi-k)_+\cdot\nabla\xi\,d\bx\\
\leq &\vep\int_{\Bbb R^d}(1+\phi^2)|\nabla(\phi-k)_+|^2\,d\bx
+C_{\vep}\int_{\Bbb R^d}(1+\phi^2)|\nabla\xi|^2(\phi-k)_+^2\,d\bx.
\end{align*}
Now choosing sufficiently small $\vep$ and defining function $\Phi_k(\bx)=(1+\phi)(\phi-k)_+$, we can get
\be
\int_{\Bbb R^d}|\nabla\Phi_k|^2|\xi|^2\,d\bx\leq C\int_{\Bbb R^d}|\nabla\xi|^2\Phi_k^2\,d\bx+C\int_{\Bbb R^d}|f|(\phi-k)_+\xi^2\,d\bx,
\ee
and
\be\label{eq:dgit1}
\int_{\Bbb R^d}|\nabla(\xi\Phi_k)|^2\,d\bx\leq C\int_{\Bbb R^d}|\nabla\xi|^2\Phi_k^2\,d\bx+C\int_{\Bbb R^d}|f|(\phi-k)_+\xi^2\,d\bx,
\ee
Since $f=-\beta|\phi|^2\phi+\mu\phi\in L^q(\Bbb R^d)$ for any $2\leq q<\infty$, we can proceed to obtain $L^\infty$ bound of $\phi$ by De Giorgi's iteration. Let $B_r(\bx)$
be the ball centered at $\bx$ with radius $r$, and we use $B_r$ for short to denote the ball centered at origin.  For $0<r<R\leq 1$, we choose $C_0^\infty$ nonnegative cutoff function $\xi(\bx)=1$ for $\bx\in B_r(\bx_0)$ and $\xi(\bx)=0$ for $\bx\notin B_R(\bx_0)$ such that
$|\nabla\xi(\bx)|\leq \frac{2}{R-r}$. Since for large $q$,
\be
\int_{\Bbb R^d}|f|(\phi-k)_+\xi^2\,d\bx\leq \|\xi f\|_{L^q}\|(\phi-k)_+\xi\|_{L^6}|\{\Phi_k\xi>0\}|^{\frac{5}{6}-\frac{1}{q}},
\ee
where $|A|$ denotes the Lebesgue measure of set $A$, we have by H\"oder inequality and Sobolev inequality in 2D and 3D, for any $\vep>0$,
\begin{align*}
&\int_{\Bbb R^d}|f|(\phi-k)_+\xi^2\,d\bx\\
\leq& C\|\xi f\|_{L^q}\|\nabla((\phi-k)_+\xi)\||\{\Phi_k\xi>0\}|^{\frac{5}{6}-\frac{1}{q}}\\
\leq& \vep \|\nabla((\phi-k)_+\xi)\|^2+C_{\vep}\|\xi f\|_{L^q}^2|\{\Phi_k\xi>0\}|^{\frac{5}{3}-\frac{2}{q}}\\
\leq& 2\vep (\|\nabla(\Phi_k\xi)\|^2+\|\Phi_k\nabla\xi\|^2)+C_{\vep}\|\xi f\|_{L^q}^2|\{\Phi_k\xi>0\}|^{\frac{5}{3}-\frac{2}{q}}.
\end{align*}
Thus, from above inequality and \eqref{eq:dgit1}, we arrive at
\be\label{eq:dgit2}
\|\nabla(\xi\Phi_k)\|^2\leq C\left(\|\Phi_k\nabla\xi\|^2+\|\xi f\|_{L^q}^2|\{\Phi_k\xi>0\}|^{\frac{5}{3}-\frac{2}{q}}\right).
\ee
Since $\xi\Phi_k\in H_0^1(B_1(\bx_0))$, we conclude by Sobolev inequality that $\forall p\in[2,6]$,
\be\label{eq:digit3}
\|\xi\Phi_k\|^2\leq \|\xi\Phi_k\|_{L^6}^2|\{\Phi_k\xi>0\}|^{1-\frac{2}{6}}\leq C(d)\|\nabla(\xi\Phi_k)\|^2|\{\Phi_k\xi>0\}|^{\frac23}.
\ee
By choosing  $q=3$, \eqref{eq:dgit2} and \eqref{eq:digit3} implies that
\be\label{eq:digit4}
\|\xi\Phi_k\|^2\leq \left(\|\Phi_k\nabla\xi\|^2|\{\Phi_k\xi>0\}|^{\frac{2}{3}}+\|f\|_{L^3(B_1(\bx_0))}^2|\{\Phi_k\xi>0\}|^{\frac{5}{3}}\right)
\ee
Denote
\be
A(k,r)=\{\bx|\bx\in B_r(\bx_0),\quad \phi>k\}.
\ee
and for $k>0$, $0<r<R\leq1$, we have
\begin{equation}\label{eq:digit5}
\int_{A(k,r)}\Phi_k^2\,d\bx\leq C(\frac{1}{(R-r)^2}|A(k,R)|^{\frac23}\int_{A(k,R)}\Phi_k^2\,d\bx+\|f\|_{L^3(B_1(\bx_0))}^2|A(k,R)|^{\frac{5}{3}}).
\end{equation}
We claim that there exists $\tilde{C}$, such that for $k=\tilde{C}(\|f\|_{L^3(B_1(\bx_0))}+\|(1+\phi)\phi\|_{L^2(B_1(\bx_0))})$,
\be\label{eq:digit-bd}
\int_{A(k,\frac12)}\Phi_k^2\,d\bx=0.
\ee
Taking $h>k>k_0$ and $0<r<1$, we find $A(h,r)\subset A(k,r)$ with
\be
\int_{A(h,r)}\Phi_h^2\,d\bx\leq\int_{A(k,r)}\Phi_k^2\,d\bx.
\ee
In addition, since $\Phi_k=(1+\phi)(\phi-k)_+$, we have
\be
|A(h,r)|=|B_r(\bx_0)\cap\{\phi-k\ge h-k\}|\leq \frac{1}{(h-k)^2}\int_{A(k,r)}\Phi_k^2\,d\bx.
\ee
Now, let us choose $\frac12\leq r<R\leq1$. We obtain from \eqref{eq:digit5} that
\begin{align*}
&\int_{A(h,r)}\Phi_h^2\,d\bx\\
\leq&C\left(\frac{1}{(R-r)^2}\int_{A(h,R)}\Phi_h^2\,d\bx+\|f\|_{L^3(B_1(\bx_0))}^2|A(h,R)|\right)|A(h,R)|^{\frac23}\\
\leq&C\left(\frac{1}{(R-r)^2}+\frac{\|f\|_{L^3(B_1(\bx_0))}^2}{(h-k)^2}\right)\frac{1}{(h-k)^{\frac43}}\left(\int_{A(k,R)}\Phi_k^2\,d\bx\right)^{\frac53},
\end{align*}
and
\be\label{eq:digit6}
\|\Phi_h\|_{L^2(B_r(\bx_0))}\leq C\left(\frac{1}{R-r}+\frac{\|f\|_{L^3(B_1(\bx_0))}}{h-k}\right)\frac{1}{(h-k)^{\frac23}}\|\Phi_k\|_{L^2(B_R(\bx_0))}^{\frac53}.
\ee
Denote function
\be
\chi(k,r)=\|\Phi_k\|_{L^2(B_r(\bx_0))}.
\ee
For some value of $k>0$ to be determined later, we define for $l=0,1,2,\ldots,$,
\be
k_l=(1-\frac{1}{2^l})k,\quad r_l=\frac12+\frac{1}{2^{l+1}},
\ee
then $k_l-k_{l-1}=\frac{k}{2^l}$ and $r_{l-1}-r_{l}=\frac{1}{2^{l+1}}$. From \eqref{eq:digit6}, we find
\begin{align*}
\chi(k_l,r_l)\leq &C\left(2^{l+1}+\frac{2^l\|f\|_{L^3(B_1(\bx_0))}}{k}\right)\frac{2^{\frac{2}{3}l}}{k^{\frac23}}(\chi(k_{l-1},r_{l-1}))^{\frac53}\\
\leq&2C\frac{\|f\|_{L^3(B_1(\bx_0))}+k}{k^{\frac{5}{3}}}2^{\frac{5}{3}l}(\chi(k_{l-1},r_{l-1}))^{\frac53}.
\end{align*}
Then, we prove that there exists $\gamma>0$ such that
\be\label{eq:ind}
\chi(k_l,r_l)\leq \frac{\chi(k_0,r_0)}{\gamma^l},\quad \gamma>1.
\ee
We argue by induction. When $l=0$, it is obvious true. Suppose \eqref{eq:ind} is true for $l-1$, i.e.
\begin{equation*}
(\chi(k_{l-1},r_{l-1}))^{\frac53}\leq   \frac{\gamma^{\frac53}(\chi(k_0,r_0))^{\frac23}}{\gamma^{\frac23l}}   \cdot  \frac{\chi(k_0,r_0)}{\gamma^{l}}.
\end{equation*}
Then, we have
\begin{align*}
\chi(k_{l},r_{l})\leq &2C\frac{\|f\|_{L^3(B_1(\bx_0))}+k}{k^{\frac{5}{3}}}2^{\frac{5}{3}l}(\chi(k_{l-1},r_{l-1}))^{\frac53}\\
\leq &2C\gamma^{\frac53}(\chi(k_0,r_0))^{\frac23}\frac{\|f\|_{L^3(B_1(\bx_0))}+k}{k^{\frac{5}{3}}}\cdot\frac{2^{\frac{5}{3}l}}{\gamma^{\frac23l}} \cdot\frac{\chi(k_0,r_0)}{\gamma^{l}}.
\end{align*}
Let us choose $\gamma>1$ such that $\gamma^{\frac23}=2^{\frac53}$. Now we want to pick $k$ sufficiently large such that
\be\label{eq:k}
2C\gamma^{\frac53}\frac{\|f\|_{L^3(B_1(\bx_0))}+k}{k}\left(\frac{\chi(k_0,r_0)}{k}\right)^{\frac23}\leq1.
\ee
Choosing $k=\tilde{C}(\|f\|_{L^3(B_1(\bx_0))}+\chi(k_0,r_0))$ for sufficiently large $\tilde{C}$, we get desired inequality \eqref{eq:k}. This gives that
\eqref{eq:ind} is true for $l$ and hence induction is done. Letting $l\to\infty$ in \eqref{eq:ind}, we find  $\chi(k,\frac12)=0$, which implies that
\be
\Phi_k(\bx)=0,\quad \forall \bx\in B_{\frac12}(\bx_0),
\ee
i.e.,
\begin{align*}
\sup_{B_{\frac12}(\bx_0)} \phi_+\leq& \tilde{C}(\|f\|_{L^3(B_1(\bx_0))}+\chi(k_0,r_0))\\
\leq& \tilde{C}(\|f\|_{L^3(B_1(\bx_0))}+\|\Phi_0\|_{L^2(B_1(\bx_0))})\\
\leq&
\tilde{C}(\|f\|_{L^3(B_1(\bx_0))}+\|\phi\|_{L^2(B_1(\bx_0))}+\|\phi\|_{L^4(B_1(\bx_0))}).
\end{align*}
The same estimates applies for $-\phi$ and we can conclude that
\begin{align*}
\|\phi\|_{L^\infty(B_{\frac12}(\bx_0))}
\leq&
\tilde{C}(\|f\|_{L^3(B_1(\bx_0))}+\|\phi\|_{L^2(B_1(\bx_0))}+\|\phi\|_{L^4(B_1(\bx_0))}).
\end{align*}
This shows $\phi$ is bounded and $\lim_{|\bx|\to0}\phi(\bx)=0$.

Thirdly, we prove that $\int_{\Bbb R^d\backslash B_R}(|\nabla\phi|^2+|\phi|^2)\,d\bx$ decays exponentially as $R\to\infty$. Choose test function
$\varphi=\eta^2(\bx)\phi$ in \eqref{eq:weak} with $\eta(\bx)$ being a smooth nonnegative cutoff function such that $\eta(\bx)=0$ for
$\bx\in B_R$ and $\eta(\bx)=1$ for $\bx\in \Bbb R^d\backslash B_{R+1}$, then the following holds
\begin{align*}
&\int_{\Bbb R^d\backslash B_R}\left((\frac12+2\delta\phi^2)|\nabla\phi|^2\eta^2+V(\bx)|\phi|^2\eta^2+\beta\phi^4\eta^2-\mu\phi^2\eta^2\right)\,d\bx\\
&=-\int_{B_{R+1}\backslash B_R} \left(1+2\delta\phi^2\right)\eta\phi\nabla\phi\cdot\nabla\eta\,d\bx.
\end{align*}
Since $\lim_{|\bx|\to\infty}V(\bx)=\infty$ and $\phi$ is bounded, we find that for large $R$,
\be
\int_{\Bbb R^d\backslash B_R}(|\phi|^2+|\nabla\phi|^2)\,d\bx\leq C\int_{B_{R+1}\backslash B_R}(|\phi|^2+|\nabla\phi|^2)\,d\bx.
\ee
Let $a_n=\int_{\Bbb R^d\backslash B_{R_n}}(|\phi|^2+|\nabla\phi|^2)\,d\bx$ with $R_n=R+n$ ($n=0,1,2,\ldots$), then
$a_{n}\leq C(a_{n+1}-a_n)$ and $a_{n+1}\leq \alpha a_n$ with $\alpha=\frac{C}{1+C}$. Hence $a_{n+1}\leq \alpha^n a_0$ which would imply the
exponential decay of $a_n$ as well as $\int_{\Bbb R^d\backslash B_R}(|\nabla\phi|^2+|\phi|^2)\,d\bx$.

Lastly, combining the exponential decay of $\int_{\Bbb R^d\backslash B_R}(|\nabla\phi|^2+|\phi|^2)\,d\bx$ and De Giorgi's iteration shown above,
we can derive the exponential fall-off of $\|\phi\|_{L^\infty}$.

(ii) The regularity of the ground state $\phi_g$ can be proved by a variable of change method \cite{Liu,ColinM}.
Let $v=F(t)$ to be the solution of the ODE $F^\prime(t)=\sqrt{\frac12+
2\delta t^2}$
with $F(0)=0$, then $F(t)$ is strictly
 increasing function, its inverse exists $t=G(v)$. Let $u=F(\phi)$, then $\phi=G(u)$ and the energy functional $E(\cdot)$ \eqref{eq:energy}
 becomes
 \be
 E(\phi)=\int_{\Bbb R^d}\left(|\nabla u|^2+V(\bx)G^2(u)+\frac{\beta}{2} G^4(u)\right)\,d\bx:=\tilde{E}(u).
 \ee
 $u_g=F(\phi_g)$ is the minimizer of $\tilde{E}(u)$ under constraint $\int_{\Bbb R^d}G(u)^2\,d\bx=1$. It follows that
 $u_g$ satisfies the following Euler-Lagrange equation (for $C_0^\infty$ test function)
 \be
 -\nabla^2 u+V(\bx)G(u)G^\prime(u)+\beta |G(u)|^2G(u)G^\prime(u)=\lambda G(u)G^\prime(u).
 \ee
 Since $\phi_g$ is bounded, we know $u_g$ is bounded, hence $G(u_g)$ and $G^\prime(u_g)$ are bounded with $\nabla ^2u_g\in L^\infty_{\rm loc}$.
 We conclude that $u_g$ is once continuously differentiable and
 $\nabla u_g$ is H\"older continuous with order 1.
  Noticing that $\nabla^2\phi_g=G^\prime(u_g)\nabla^2 u_g+G^{\prime\prime}(u_g)|\nabla u_g|^2$,
 we find  that $\phi_g$ is once continuously differentiable and
 $\nabla \phi_g$ is H\"older continuous with order 1.
  In addition, if $V\in C^\infty$, we can obtain  $\phi_g\in C^\infty$ by a bootstrap argument using the $L^\infty$ bound of $\phi_g$. $\hfill\Box$

\section{Limiting behavior of ground states in the whole space}\label{subsec:ws}
In this section, we consider the behavior of the ground state \eqref{eq:minimize} in different $\beta$ and $\delta$ parameter regimes for typical potentials in the whole space, e.g. harmonic potential.
In the next section, we will discuss about the box potential case (typical potential in bounded domains).

For the whole space, the harmonic trapping potential  \eqref{eq:potential:ho}   is the most relevant experimental case and we will focus on such potentials.
Our results are valid for more general confining potentials
in $d$ ($d=1,2,3$) dimensions, e.g. confining potentials satisfying the homogeneous conditions $V(\lambda\bx)=|\lambda|^sV(\bx)$, for some $s>0$ and any $\lambda\in\mathbb{R}$.

There are two interesting parameter regimes including the large $\beta,\delta$ limit and the vanishing $\delta$ limit. Below, we start with the strong interaction regime, i.e. $\delta\gg1$ and $|\beta|\gg1$ ($\delta\ge0$ is necessary for the existence of the ground state).
\subsection{Thomas-Femi (TF) limit}
When the number of particles in BEC is relatively large, the interaction between particles is dominant while the kinetic energy term can be neglected, leading to the Thomas-Fermi (TF) limit/approximation. The TF limit for the GPE has been thoroughly studied in \cite{Bao2013,LiebL,Lie} while the generalization to the MGPE is missing. As shown in \cite{Ruan}, the TF limit for the MGPE is more complicated and more interesting where new phenomenons can be observed due to the competition between the cubic nonlinear term and the new one.
In this section, we aim to give a rigorous mathematical characterization of the TF approximations in different parameter regimes under a specific external potential, i.e. the harmonic potential.

When $V(\bx)$ is the harmonic potential \eqref{eq:potential:ho},  we consider the limiting profile of ground states \eqref{eq:minimize} under different sets of parameters $\delta\gg1$
and $|\beta|\gg1$.
For any $\phi(\bx)\in S$, choose proper scaling $\phi^\vep(\bx)=\vep^{-d/2}\phi(\bx/\vep)\in S$ where $\vep^{-1}$ is the length scale for the condensate width, i.e.
\be\label{eq:phieps}
\phi(\bx)=\vep^{d/2}\phi^\vep(\bx\vep),\ee we find the energy $E(\cdot)$ \eqref{eq:energy} can be written as
\begin{align}
E(\phi)=&\int_{\Bbb R^d}\left[\frac{\vep^2}{2}|\nabla\phi^\vep|^2+\frac{1}{\vep^2}V(\bx)|\phi^\vep|^2+\frac{\beta\vep^d}{2}|\phi^\vep|^4+\frac{\delta\vep^{2+d}}{2}\left|\nabla|\phi^\vep|^2\right|^2\right]\,d\bx
\nonumber\\
=&\,\frac{1}{\vep^2}\int_{\Bbb R^d}\left[\frac{\vep^{4}}{2}|\nabla\phi^\vep|^2+V(\bx)|\phi^\vep|^2+\frac{\beta\vep^{d+2}}{2}|\phi^\vep|^4+\frac{\delta\vep^{4+d}}{2}\left|\nabla|\phi^\vep|^2\right|^2\right]\,d\bx,
\end{align}
which indicates that the ground state of \eqref{eq:minimize} is equivalent to the ground state of the following energy functional $E^\vep(\cdot)$ under the constraint $\phi^\vep\in S$ through relation \eqref{eq:phieps},
\be\label{eq:energyeps}
E^\vep(\phi^\vep)=\int_{\Bbb R^d}\left[\frac{\vep^{4}}{2}|\nabla\phi^\vep|^2+V(\bx)|\phi^\vep|^2+\frac{\beta\vep^{d+2}}{2}|\phi^\vep|^4+\frac{\delta\vep^{4+d}}{2}\left|\nabla|\phi^\vep|^2\right|^2\right]\,d\bx.
\ee
Now, we give the characterization of the ground state $\phi_g$ for \eqref{eq:minimize} when the two interaction strengths $|\beta|$ and $\delta>0$ are very large.
By comparing the contributions from each term in the energy \eqref{eq:energyeps} \cite{Ruan}, we can identify the following different regimes:

 Case 1: $\beta\to +\infty$ and $\delta/\beta^{\frac{4+d}{2+d}}\ll1$, i.e. $\delta=o(\beta^{\frac{4+d}{2+d}})$;\\
 \hspace*{.4cm} Case 2: $\beta\to+\infty$ and $\lim_{\beta\to+\infty}\delta/\beta^{\frac{4+d}{2+d}}=\delta_\infty>0$;\\
 \hspace*{.4cm} Case 3: $\beta\to+\infty$ and  $\delta/\beta^{\frac{4+d}{2+d}}\gg1$, i.e. $\beta=o(\delta^{\frac{2+d}{4+d}})$ as $\delta\to+\infty$;

\noindent and for $\beta\to-\infty$

 Case $1^\prime$: $\beta\to -\infty$ and $\delta/|\beta|^{\frac{4+d}{2+d}}\ll1$, i.e. $\delta=o(|\beta|^{\frac{4+d}{2+d}})$;\\
 \hspace*{.4cm} Case $2^\prime$: $\beta\to-\infty$ and $\lim_{\beta\to-\infty}\delta/|\beta|^{\frac{4+d}{2+d}}=\delta_\infty>0$;\\
 \hspace*{.4cm} Case $3^\prime$: $\beta\to-\infty$ and  $\delta/|\beta|^{\frac{4+d}{2+d}}\gg1$, i.e. $|\beta|=o(\delta^{\frac{2+d}{4+d}})$ as $\delta\to+\infty$.

Intuitively, for cases 1 and $1^\prime$, the $\beta$ cubic term in \eqref{eq:energyeps} is dominant and determines the length scale $\vep^{-1}$ of the ground state, which
increases ($\vep\to0^+$) for growing repulsive interactions $\beta\to+\infty$.   For growing attractive interactions $\beta\to-\infty$ in case 1$^\prime$, it is readily to check
that only the cubic $\beta$ term (attractive) and the $\delta$ term (repulsive) are important, where the length scale $\vep$ could go to 0 or $+\infty$ (see Theorem \ref{thm:tfneg}). For cases
3 and $3^\prime$, the $\delta$ higher order interaction term in \eqref{eq:energyeps} is dominant and the length scale $\vep\to0^+$ as $\delta\to+\infty$. For cases 2 and $2^\prime$,
the $\delta$ higher order interaction term and the $\beta$ cubic term are comparable and the length scale $\vep\to0^+$.
As shown later, the energy $E^\vep$ \eqref{eq:energyeps} with proper choice of $\vep$ ($\vep\to0^+$ for cases 1,2,3, $2^\prime$ and $3^\prime$; $\vep\to+\infty$ for case $1^\prime$ and consider $E^\vep/(\beta\vep^{d+2})$) converges to the following functionals,
\begin{align}
\label{eq:es}
&E_1(\phi)=\int_{\Bbb R^d}\left(V(\bx)|\phi|^2+\frac{1}{2}|\phi|^4\right)\,d\bx,\quad \text{for case 1},\\
\label{eq:es2}
&E_2(\phi)=\int_{\Bbb R^d}\left(V(\bx)|\phi|^2+\frac{|\phi|^4}{2}
+\frac{\delta_\infty}{2}|\nabla|\phi|^2|^2\right)\,d\bx,\quad \text{for case 2},\\
\label{eq:es1}
&E_3(\phi)=\int_{\Bbb R^d}\left(V(\bx)|\phi|^2+\frac{1}{2}|\nabla|\phi|^2|^2\right)\,d\bx, \quad\text{for case 3 and } 3^\prime,\\
\label{eq:edelta}
&E_{2^\prime}(\phi)=\int_{\Bbb R^d}\left(V(\mathbf{x})|\phi|^2-\frac{1}{2}|\phi|^4+\frac{\delta_\infty}{2}\left|\nabla|\phi|^2\right|^2\right)\,d\bx,\quad
\text{for case } 2^\prime,\\
\label{eq:eeta}
&E_{1^\prime}(\phi)=\int_{\Bbb R^d}\left(\frac{1}{2}\left|\nabla|\phi|^2\right|^2-\frac{1}{2}|\phi|^4\right)\,d\bx, \quad\text{for case } 1^\prime,
\end{align}
under the constraint
\be\label{eq:constraint}
\|\phi\|=1.
\ee The limiting profiles of the ground state \eqref{eq:minimize} can be proved to be the minimizers of the above energy functionals
\eqref{eq:es}-\eqref{eq:eeta} with constraint $\|\phi\|=1$ in different cases under proper scaling factor $\vep$.

First of all, we investigate some basic properties of the limiting profiles, i.e. the minimizer of the energy functionals \eqref{eq:es}-\eqref{eq:eeta} under \eqref{eq:constraint}.
\begin{theorem}\label{thm:fb}(Properties of the limiting profiles) Assume $0\leq V(\bx)\in L_{\rm loc}^\infty(\Bbb R^d)$  ($d=1,2,3$) satisfies $\lim\limits_{|\bx|\to\infty}V(\bx)\to+\infty$,  for each energy functional \eqref{eq:es}-\eqref{eq:eeta} with constraint \eqref{eq:constraint}, there exists a nonnegative minimizer $\sqrt{\rho_g}=\phi_g$ with $\|\phi_g\|=1$ and such nonnegative minimizer is
unique for \eqref{eq:es}, \eqref{eq:es2} and \eqref{eq:es1}. We denote the density $\rho_g=|\phi_g|^2$ and $\rho_g\ge0$ with $\|\rho_g\|_{L^1}=1$. The following properties for
the minimizer $\phi_g$ (or density $\rho_g$) hold.

(1) For \eqref{eq:es}, the density $\rho_g$  is given by $\rho_g=\max\{\mu-V(\bx),0\}$ with $\mu=E_1(\sqrt{\rho_g})+\frac12\|\rho_g\|_{L^2}^2$ and $\|\rho_g\|_{L^1}=1$.

(2) For \eqref{eq:es2}, $\rho_g\in C_{\rm loc}^{1,\alpha}\subset W_{\rm loc}^{2,p}$ ($1<p<\infty$ and $\alpha<1$) solves the free boundary value problems
\begin{align}\label{eq:fb:1}
&-\delta_\infty\Delta\rho_g+\rho_g=\left(\mu-V(\bx)\right)\chi_{\{\rho_g>0\}},
\end{align}
where $\mu=2E_2(\sqrt{\rho_g})-\int_{\Bbb R^d}V(\bx)\rho_gd\bx$.  The  conditions at the free boundaries are
\be\label{eq:fbbc}
\rho_g|_{\partial\{\rho_g>0\}}=0,\quad |\nabla\rho_g||_{\partial\{\rho_g>0\}}=0.
\ee
If $V(\bx)$ is radially increasing, we have that $\rho_g(\bx)$ is  radially decreasing and compactly supported.

(3) For \eqref{eq:es1},  $\rho_g\in C_{\rm loc}^{1,\alpha}\subset W_{\rm loc}^{2,p}$ ($1<p<\infty$ and $\alpha<1$) solves the free boundary value problems
\be
\label{eq:fb:2}
-\delta_\infty\Delta\rho_g=\left(\mu-V(\bx)\right)\chi_{\{\rho_g>0\}},
\ee
where $\mu=2E_3(\sqrt{\rho_g})-\int_{\Bbb R^d}V(\bx)\rho_g\,d\bx$. The conditions at the free boundaries are given as \eqref{eq:fbbc}. If $V(\bx)$ is radially  increasing, $\rho_g(\bx)$ is  radially decreasing and compactly supported.

(4) For \eqref{eq:eeta},  there exists a non-increasing radially symmetric minimizer $\phi_g$ which is  unique and compactly supported. The density $\rho_g=|\phi_g|^2\in C_{\rm loc}^{1,\alpha}\subset W_{\rm loc}^{2,p}$ ($1<p<\infty$ and $\alpha<1$).
In fact, $\rho_\infty$ solves the equation
\be
-\Delta \rho_g-\rho_g=\mu\chi_{\{\rho_g>0\}},\quad \mu=2E_{1^\prime}(\sqrt{\rho_g}).
\ee
\end{theorem}
 {\noindent\it Proof}. The existence and uniqueness results (except for \eqref{eq:eeta}) are straightforward following Theorem \ref{thm:mres} and the conventional GPE case \cite{Lie,Bao2013}. We omit the details here and the proof for \eqref{eq:eeta} will be shown below.

 (1) It is the classical TF density \cite{Lie}.

 (2) We first show \eqref{eq:fb:1} is valid. We adapt an approach for the classical obstacle problem in \cite{Petro}.
Since $\rho_g\ge0$ minimizes $E_2(\sqrt{\rho})$
under the constraints $\|\rho\|_{L^1}=1$ and $\rho\ge0$, in addition $V(\bx)\ge0$, we can conclude that $\rho_g$ minimizes the
following energy
\be\label{eq:es2t}
\tilde{E}(\rho)=\int_{\Bbb R^d}\left(V(\bx)|\rho|+\frac{\rho^2}{2}
+\frac{\delta_\infty}{2}|\nabla\rho|^2\right)\,d\bx,\quad \int_{\Bbb R^d}\rho(\bx)\,d\bx=1,
\ee
i.e.  $\rho_g$ is still a minimizer if we remove the nonnegative constraint with the price to have a non-smooth $V(\bx)|\rho|$ term. The reason is that if
$\int_{\Bbb R^d}\rho(\bx)\,d\bx=1$, we can write $\rho_+(\bx)=\max\{\rho(\bx),0\}$ and $\rho_-(\bx)=\max\{-\rho(\bx),0\}$, and $\int_{\Bbb R^d}\rho_+(\bx)\ge1$. Since all
the terms in energy $\tilde{E}(\rho)$ are positive, we have $\tilde{E}(\rho_+/\|\rho_+\|_{L^1})\leq \tilde{E}(\rho_+)\leq\tilde{E}(\rho)$. Thus, the minimizer must be nonnegative
and the unique minimizer
of \eqref{eq:es2t} (by convexity) is $\rho_g$.

Now, we would like to derive the equation for $\rho_g$. In order to do this, we introduce the following regularization of \eqref{eq:es2t}. Mollify
 the step function $\chi_{[0,\infty)}(s)$ ($s\in\Bbb R$) to get smooth function $g_\vep(s)\in C^\infty(\Bbb R)$ ($\vep>0$) such that $g_\vep(s)=1$ if $s>0$, $g_\vep(s)=0$ if
 $s\leq-\vep$ and $g_\vep^\prime(s)\ge 0$ for all $s\in\Bbb R$. Moreover, $g_\vep(s)\to \chi_{(0,\infty)}$ as $\vep\to0^+$. Denote $G_\vep(s)=\int_{-\infty}^sg_\vep(s)\,ds$ and
 $G^{\prime\prime}\ge0$ indicating that $G_{\vep}$ is a convex function.
Now, let us consider
\be\label{eq:es2t2}
\tilde{E}^\vep(\rho)=\int_{\Bbb R^d}\left(V(\bx)G_\vep(\rho(\bx))+\frac{\rho^2}{2}
+\frac{\delta_\infty}{2}|\nabla\rho|^2\right)\,d\bx,\quad \int_{\Bbb R^d}\rho(\bx)\,d\bx=1,
\ee
which is still a convex minimization problem and we have a unique  minimizer $\rho^{\vep}_g(\bx)\ge0$. Moreover, we can find the equations for $\rho^{\vep}_g(\bx)$.

For any compactly supported smooth function $\varphi\in C_c^\infty(\Bbb R^d)$, consider $h(s)=\tilde{E}^{\vep}(\rho_s(\bx))$ where
$\rho_s(\bx)=(\rho^\vep_g+s\varphi)/\int_{\Bbb R^d}(\rho^\vep_g+s\varphi)\,d\bx$ and $s\in(-s_0,s_0)$ with sufficiently small $s_0>0$ such that
$\int_{\Bbb R^d}(\rho^\vep_g+s\varphi)\,d\bx\ge 1/2$,
we then have $h(s)$ attains its minimum at $s=0$. By standard computations and arguments  \cite{Evans,LiebL}, we can get that there exists
a Lagrangian multiplier $\mu_\vep$, such that $\rho^\vep_g$ solves (in the weak sense)
\begin{equation}\label{eq:el-1}
-\delta_\infty\Delta\rho^\vep_g+\rho^\vep_g=\mu^\vep-V(\bx)g_\vep(\rho^\vep_g).
\end{equation}
It is easy to see that $\mu^\vep$ is uniformly bounded and $\mu^\vep-V(\bx)g_\vep(\rho^\vep_g)\in L_{\rm loc}^\infty$, which implies that  for
any bounded smooth domain $\Omega\subset\Bbb R^d$, $\rho_g^\vep$ is uniformly bounded in $W^{2,p}(\Omega)$  ($p\in(1,p)$) by classical elliptic regularity results \cite{Evans,LiebL}.
Using Sobolev embedding,
 $\rho_g^\vep$ is uniformly bounded in $C^{1,\alpha}(\Omega)$ (for some $0<\alpha<1$) locally and hence there exist $\tilde{\rho}\in W^{2,p}(\Omega)$ such that as $\vep\to0^+$ (take a subsequence
  $\vep_k\to0^+$ if necessary),
 $\rho_g^\vep$ converges to $\tilde{\rho}$ strongly in $C_{\rm loc}^{1,\alpha}$ and weakly in $W^{2,p}_{\rm loc}$. Consequently, $\tilde{\rho}\ge0$
 and $\|\tilde{\rho}\|_{L^1}=1$ ($V(\bx)$ is a confining potential).
 In fact, we can show $\tilde{\rho}=\rho_g$. Passing to the limit as $\vep\to0^+$ in $\tilde{E}(\rho_g^\vep)\leq\tilde{E}^\vep(\rho_g^\vep)\leq\tilde{E}^\vep(\rho_g)$ ($G_\vep(|s|)\ge |s|$),
 we  observe that
 $\tilde{E}(\tilde{\rho})\leq \limsup\limits_{\vep\to0^+}\tilde{E}^\vep(\rho_g^\vep)\leq\tilde{E}(\rho_g)$ and it is obvious $\tilde{\rho}=\rho_g$.

Now, we have $\rho_g\in W^{2,p}_{\rm loc}\cap C_{\rm loc}^{1,\alpha}$ and we want to show that
 \begin{equation}\label{eq:el-2}
-\delta_\infty\Delta\rho_g+\rho_g=(\mu-V(\bx))\chi_{\{\rho_g>0\}},\quad \text{a.e.} \quad\bx\in \Bbb R^d.
\end{equation}
Since $\rho^\vep_g\in W_{\rm loc}^{2,p}$ is a strong solution of \eqref{eq:el-1}, \eqref{eq:el-1} is valid almost everywhere. In addition, $\rho^\vep_g\to \rho_g$ in $C_{\rm loc}^{1,\alpha}$,
so we can pass to the limit as $\vep\to0^+$ in \eqref{eq:el-1} to get
\be
-\delta_\infty\Delta\rho_g+\rho_g=\mu-V(\bx),\quad \text{a.e.}\quad\bx\in\{\rho_g>0\},
\ee
where $\mu$ is a limiting point of $\mu_\vep$ as $\vep\to0^+$ (take a subsequence if necessary here).
On the other hand, $\rho_g\in W^{2,p}_{\rm loc}$ implies $\Delta \rho_g=0$ a.e. $\bx\in\{\rho_g=0\}$. Together, we have shown $\rho_g$ is the solution of the free boundary value problem
\eqref{eq:fb:1} and $\mu$ can be computed via multiplying both sides of \eqref{eq:fb:1} by $\rho_g$ and integrating over $\Bbb R^d$.

 When $V(\bx)=V(r)$ ($r=|\bx|$) is radially increasing, it is easy to find $\rho_j(\bx)$ is radially decreasing  by Schwarz rearrangement \cite{LiebL}. Here we would like to show such ground state is compactly supported.
For simplicity, we write $\rho_g(\bx)=\rho_g(|\bx|)=\rho_g(r)$ ($r=|\bx|$) and $\rho_g^\prime(r)\leq0$. Integrating \eqref{eq:fb:1} over ball $B_R=\{|\bx|< R\}$, we get
\begin{equation*}
\int_{B_R} \left(V(\bx)\chi_{\{\rho_g>0\}}+\rho_g(\bx)\right)\,d\bx-\delta_{\infty}\int_{\partial B_R}\partial_{n}\rho_g(\bx)dS=\mu\int_{B_R}\chi_{\{\rho_g>0\}}\,d\bx,
\end{equation*}
where $\partial_{n}\rho_g(\bx)|_{\partial B_R}\leq0$ ($\rho_g(r)$ is non-increasing). On the other hand, $\lim\limits_{r\to\infty}V(r)=\infty$, choosing $R_0$ large enough such that
$V(r)\ge2\mu$ ($r\ge R_0$),  we have
\begin{equation*}
\int_{B_R \setminus B_{R_0}} \left[(V(\bx)-\mu)\chi_{\{\rho_g>0\}}+\rho_g(\bx)\right]\,d\bx\leq \mu |B_{R_0}|,
\end{equation*}
which is true for all $R>0$. Thus, we arrive at
\be
|B_{R_0}^c\cap \chi_{\{\rho_g>0\}}|\leq|B_{R_0}|,
\ee
and it implies that $|\{\rho_g>0\}|<\infty$. Therefore $\rho_g$ is compactly supported.

(3) The proof is quite similar to the proof of item (2) and is omitted for brevity.

(4) We show the fact that the decreasing radially symmetric minimizer $\rho_\infty$ of \eqref{eq:eeta} exists and is unique.  In view of Nash inequality,
$E_{1^\prime}(\sqrt{\rho})$ is bounded from below
under constraint $\|\rho\|_{L^1}=1$ with $\rho\ge0$. By Schwarz rearrangement, we can take a minimizing sequence of non-increasing radially symmetric functions
 $\{\rho_n\}_{n=1}^\infty$ where $\|\rho_n\|_{L^1}=1$ and $\|\rho_n\|_{H^1}\leq C$. Therefore, there exists
 $\rho_\infty\in H^1$ such that a subsequence (denoted as the original sequence) $\rho_n\to \rho_\infty$ weakly in $H^1$.  
   In addition, for the non-increasing radially symmetric function
 $\rho_n$,
 \be
 |\rho_n(\bx)|\leq \frac{C}{R^d}\|\rho_n\|_{L^1}\leq\frac{C}{R^d},\quad |\bx|\ge R>0,
 \ee
 which would imply $\rho_n\to \rho_\infty$ strongly in $L^2$ and so $\rho_\infty\ge0$ with $\|\rho_\infty\|_{L^1}\leq1$ . In fact, we can show $\|\rho_\infty\|_{L^1}=1$. Denote
 $I_\alpha=\inf_{\rho\ge0,\|\rho\|_{L^1}=\alpha} E_{1^\prime}(\rho)$ ($\alpha>0$), then it is obvious $I_\alpha=\alpha^2I_1$.
 For any $\rho\ge0$ with $\|\rho\|_{L^1}=1$, denote $\rho_{\eta}=\eta^{-d}\rho(\bx/\eta)$ ($\eta>0$) and
 we have $\|\rho_\eta\|_{L^1}=1$ and $E_{1^\prime}(\rho_\eta)=\frac{1}{2\eta^{d+2}}\|\nabla\rho\|^2-\frac{1}{2\eta^{d}}\|\rho\|^2<0$  for $\eta>\|\nabla\rho\|/\|\rho\|$,
 which immediately suggests that $I_1<0$. If $\|\rho_\infty\|_{L^1}=\alpha<1$, by the
 convergence of $\rho_n$, we get
 \be
 \alpha^2 I_1=I_\alpha\leq E_{1^\prime}(\sqrt{\rho_\infty})\leq \liminf\limits_{n\to+\infty}E_{1^\prime}(\sqrt{\rho_n})=I_1,
 \ee
 which leads to $I_1\ge0$ contradicting to the fact $I_1<0$. Thus $\|\rho_\infty\|_{L^1}=1$ and $\rho_\infty$ is a non-increasing radially symmetric minimizer of \eqref{eq:eeta}.
 Next, we show such minimizer is unique. Following Theorem \ref{thm:fb}, we can get the equations for the minimizers of \eqref{eq:eeta} as
 \be\label{eq:fb:3}
 -\Delta \rho-\rho=\mu\chi_{\{\rho>0\}},
 \ee
 and a  non-increasing radially symmetric minimizer  $\rho$ is compactly supported with the regularity stated in Theorem \ref{thm:fb}.
 If there are two non-increasing radially symmetric minimizers $\rho_1$ and $\rho_2$
 to the energy \eqref{eq:eeta}, we have
  \begin{equation*}
 -\Delta \rho_1-\rho_1=\mu_1\chi_{\{\rho_1>0\}},\quad  -\Delta \rho_2-\rho_2=\mu_2\chi_{\{\rho_2>0\}},
 \end{equation*}
 and $\mu_1=\mu_2=2I_1$ (multiplying both sides of the equation by $\rho_j$, j=1,2, and integrate). Thus, by integrating the equations, we know $\rho_1$ and $\rho_2$ have the same supports (denote as the ball $B_R$). $\rho_1=\rho_2$
 is then a consequence of classical ODE theory by noticing that $\rho_j(R)=\partial_r\rho_j(R)=0$. The existence and uniqueness of
 non-increasing radially symmetric minimizers are proved. $\hfill\Box$

   Below, we specify the convergence towards the limiting profiles in Theorem \ref{thm:fb} in different cases.
\begin{theorem}\label{thm:TFp}(positive $\beta$ limit) Let $V(\bx)$ ($\bx\in\Bbb R^d$, $d=1,2,3$) be given in \eqref{eq:potential:ho}, $\delta>0$, $\phi_g\in S$ be the positive ground state of \eqref{eq:minimize},
and $\phi_g^\vep(\bx)=\vep^{-d/2}\phi_g(\bx/\vep)\in S$ for some $\vep>0$ depending on $\beta$ and $\delta$.

(1) For Case 1, i.e. $\beta\to +\infty$ and $\delta/\beta^{\frac{4+d}{2+d}}\ll1$, set 
$\vep=\beta^{-\frac{1}{2+d}}$. For $\beta\to+\infty$ ($\vep\to0^+$),  we have
\be
\rho_g^\vep=|\phi_g^\vep(\bx)|^2\rightarrow\rho_\infty(\bx)=|\phi_\infty(\bx)|^2 \text{ in } L^2,
\ee
where $\phi_\infty(\bx)$ is the unique nonnegative minimizer of the energy \eqref{eq:es}.

(2) For Case 2, i.e. $\beta\to+\infty$ and 
$\lim_{\beta\to+\infty}
\delta/\beta^{\frac{4+d}{2+d}}=\delta_\infty>0$,
set 
$\vep=\beta^{-\frac{1}{2+d}}$. For $\delta\to+\infty$ ($\vep\to 0^+$), we have
\be
\rho_g^\vep(\bx)=|\phi_g^\vep(\bx)|^2 \rightarrow \rho_\infty(\bx)=|\phi_\infty(\bx)|^2 \text{ in } H^1,
\ee
where $\phi_\infty(\bx)$ is the unique nonnegative minimizer of the energy \eqref{eq:es2}.

(3) For Case 3, i.e. $\beta\to+\infty$ and $\delta/\beta^{\frac{4+d}{2+d}}\gg1$,  set
$\vep=\delta^{-\frac{1}{4+d}}$. For $\delta\to+\infty$ ($\vep\to 0^+$), we have
\be
\rho_g^\vep(\bx)=|\phi_g^\vep(\bx)|^2\rightarrow\rho_\infty(\bx)=|\phi_\infty(\bx)|^2 \text{ in } H^1,
\ee
where $\phi_\infty(\bx)$ 
 is the unique nonnegative minimizer of the energy \eqref{eq:es1}.

\end{theorem}
{\noindent\it Proof}. We separate the three cases.

 (1) Using \eqref{eq:energyeps} and choosing $\vep=\beta^{-1/(d+2)}$, we find  $\phi_g^\vep\in S$ minimizes
\be
E^\vep(\phi^\vep)=\int_{\Bbb R^d}\left[\frac{\vep^4}{2}|\nabla\phi^\vep|^2+V(\bx)|\phi^\vep|^2+\frac{|\phi^\vep|^4}{2}
+\frac{\delta\vep^{4+d}}{2}\left|\nabla|\phi^\vep|^2\right|^2\,d\bx\right].
\ee
On the other hand, $E_1(\phi)$ has a unique nonnegative minimizer $\phi_\infty$ and by an approximation argument, we can take any smooth approximation of $\phi_\infty(\bx)$ in $S$ and find that for any $\eta>0$ with  $\delta=o(\beta^{\frac{d+4}{d+2}})$
\begin{equation*}
E_1(\phi_\infty)\leq  E^{\vep}(\phi_g^\vep) \leq E_1(\phi_\infty)+\eta+C(\eta)(\vep^{4}+o(1)),
\end{equation*}
which implies
\be
\lim_{\vep\to0^+}E_1(\phi^\vep_g)=E_1(\phi_\infty).
\ee
Hence we know $\phi^\vep_g$ ($\vep\to0^+$) is  a minimizing sequence for $E_1(\cdot)$. On the other hand,
\begin{align*}
E_1(\phi^\vep_g)-E_1(\phi_\infty)=&\int_{\Bbb R^d}\left[(V(\bx)+|\phi_\infty|^2)(|\phi_g^\vep|^2-|\phi_\infty|^2) +\frac{1}{2}(|\phi_g^\vep|^2-|\phi_\infty|^2)^2\right]\,d\bx\\
=&\int_{\Bbb R^d}\left[\max\{V(\bx),\mu\}(|\phi_g^\vep|^2-|\phi_\infty|^2) +\frac{1}{2}(|\phi_g^\vep|^2-|\phi_\infty|^2)^2\right]\,d\bx\\
\ge&\int_{\Bbb R^d}\left[\mu|\phi_g^\vep|^2-\mu|\phi_\infty|^2 +\frac{1}{2}(|\phi_g^\vep|^2-|\phi_\infty|^2)^2\right]\,d\bx\\
=&\frac{1}{2}\int_{\Bbb R^d}(|\phi_g^\vep|^2-|\phi_\infty|^2)^2d\bx,
\end{align*}
and the conclusion follows.

(2)
Similar to the part (1),  it is easy to show $\lim\limits_{\vep\to0^+}E_2(\phi^\vep_g)=E_2(\phi_\infty)$.
Noticing that  for any function $0\leq \sqrt{\rho(\bx)}\in H^1$ with $\int_{\Bbb R^d}\rho(\bx)=1$, we have $E_2(\sqrt{(\rho_\infty+s\rho)/(1+s)})$ ($s\ge0$) attains
minimum at $s=0$.
By direct computation, we find
\begin{align*}
\frac{d}{ds}E_2\left(\sqrt{\frac{\rho_\infty+s\rho}{1+s}}\right)\bigg|_{s=0}=&\int_{\Bbb R^d} (V(\bx)\rho(\bx)+\rho_\infty(\bx)\rho(\bx)+\delta_\infty \nabla\rho_\infty(\bx)\cdot\nabla\rho(\bx))\,d\bx\\
&-\int_{\Bbb R^d} (V(\bx)\rho_\infty(\bx)+\rho^2_\infty(\bx)+\delta_\infty |\nabla\rho_\infty(\bx)|^2)\,d\bx\\
\ge&0.
\end{align*}
A simple calculation shows
\begin{align*}
&E_2(\phi^\vep_g)-E_2(\phi_\infty)\\
=&\frac{d}{ds}E_2\left(\sqrt{\frac{\rho_\infty+s\rho^\vep}{1+s}}\right)\bigg|_{s=0}+\int_{\Bbb R^d}\left[\frac{1}{2}(|\phi_g^\vep|^2-|\phi_\infty|^2)^2
+\frac{\delta_\infty}{2}(\nabla|\phi_g^\vep|^2-\nabla|\phi_\infty|^2)^2\right]\,d\bx\\
\ge&\int_{\Bbb R^d}\left[\frac{1}{2}(|\phi_g^\vep|^2-|\phi_\infty|^2)^2
+\frac{\delta_\infty}{2}(\nabla|\phi_g^\vep|^2-\nabla|\phi_\infty|^2)^2\right]\,d\bx,
\end{align*}
which implies $\rho_g^\vep(\bx)=|\phi_g^\vep(\bx)|^2$ converges to $\rho_\infty(\bx)$ in $H^1$.

(3)
Using \eqref{eq:energyeps} and choosing $\vep=\delta^{-\frac{1}{4+d}}$, we find  $\phi_g^\vep\in S$ minimizes
\be
E^\vep(\phi^\vep)=\int_{\Bbb R^d}\left[\frac{\vep^4}{2}|\nabla\phi^\vep|^2+V(\bx)|\phi^\vep|^2+\frac{\beta\vep^{d+2}|\phi^\vep|^4}{2}+\frac{1}{2}\left|\nabla|\phi^\vep|^2\right|^2\right]\,d\bx.
\ee
Nash inequality and Young inequality imply that for $\rho^\vep=|\phi^\vep|^2$ ($\phi^\vep\in S$),
\begin{align*}
\int_{\Bbb R^d}|\rho^\vep(\bx)|^2\,d\bx\leq C\|\rho^\vep\|_{L^1}^{4/d+2}\|\nabla\rho^\vep\|^{2d/d+2}
\leq  C+\|\nabla\rho^\vep\|^2.
\end{align*}
Thus, we conclude that for $\beta=o(\delta^{\frac{2+d}{4+d}})$,
\be\label{eq:1sd}
E^{\vep}(\phi^\vep)\ge \int_{\Bbb R^d}\left(V(\bx)|\phi^\vep|^2+\frac12(1-o(1))|\nabla|\phi^\vep|^2|^2\right)\,d\bx-o(1),\quad
\phi^\vep\in S.
\ee
For sufficient small $\vep$, \eqref{eq:1sd} gives that for the ground state $\phi_g^\vep$,
\be
E_3(\phi_g^\vep)\leq C,
\ee
and we obtain
\be\label{eq:2sd}
E^{\vep}(\phi_g^\vep)\ge E_3(\phi_g^\vep)-o(1).
\ee
Choosing  smooth approximations of $\phi_\infty$ in $S$ if necessary, we  could get for any $\eta>0$,
\be\label{eq:3sd}
E^{\vep}(\phi_g^\vep)\leq E_3(\phi_\infty)+\eta+ C(\eta)(\vep^{4}+o(1)).
\ee
Combining \eqref{eq:2sd}, \eqref{eq:3sd} and the fact that $\phi_\infty$ minimizes $E_3$ under the constraint $\|\phi\|=1$, we find that
\be
\lim\limits_{\vep\to0^+} E_3(\phi_g^\vep)=E_3(\phi_\infty).
\ee
On the other hand,  $E_3(\sqrt{(\rho_\infty+s\rho_g^\vep)/(1+s)})$ ($s\ge0$) reach its minimum at $s=0$, and
\begin{align*}
0\leq\frac{d}{ds}E_3\left(\sqrt{\frac{\rho_\infty+s\rho_g^\vep}{1+s}}\right)\bigg|_{s=0}
=&\int_{\Bbb R^d}\left(\rho_g^\vep V(\bx)+\nabla \rho_g^\vep\cdot\nabla \rho_\infty\right)\,d\bx\\
&-\int_{\Bbb R^d}\left(\rho_\infty V(\bx)+\nabla \rho_\infty\cdot\nabla \rho_\infty\right)\,d\bx.
\end{align*}
Therefore.
\begin{align*}
 &E_3(\phi_g^\vep)-E_3(\phi_\infty)\\
 =& \int_{\Bbb R^d}\left((\rho_g^\vep-\rho_\infty)V+\nabla(\rho_g^\vep-\rho_\infty)\cdot\nabla\rho_\infty\right)\,d\bx+\frac12\|\nabla\rho_g^\vep-\nabla\rho_\infty\|^2\\
 \ge&\frac12\|\nabla\rho_g^\vep-\nabla\rho_\infty\|^2.
\end{align*}
The convergence of $\rho_g^\vep$ towards $\rho_\infty$ as $\vep\to0^+$ is then a direct consequence. $\hfill\Box$

Theorem \ref{thm:TFp} concerns about the case $\beta>0$. However, as shown in Theorem \ref{thm:mres}, the ground state exists for negative $\beta$ as long as $\delta$ is positive.
Now we consider the interesting cases when $\beta\to-\infty$ with $\delta\to+\infty$ in different ways.
\begin{theorem}\label{thm:tfneg}
(Negative $\beta$ limit) Let $V(\bx)$ ($\bx\in\Bbb R^d$, $d=1,2,3$) be given in \eqref{eq:potential:ho}, $\beta<0$, $\delta>0$, $\phi_g\in S$
be a nonnegative ground state of \eqref{eq:minimize} and $\phi_g^\vep(\bx)=\vep^{-d/2}\phi_g(\bx/\vep)\in S$ for some $\vep>0$ depending on $\beta$ and $\delta$.

(1) For case $1^\prime$, i.e. $\beta\to-\infty$, $\delta=o(|\beta|^{\frac{4+d}{2+d}})$ as $|\beta|\to\infty$,
let 
$\varepsilon=|\beta|^{1/2}/\delta^{1/2}$, and it can be checked that, when $\beta\to-\infty$, $\varepsilon\gg1$ if $\delta=o(\beta)$ and $\varepsilon\ll1$ if $|\beta|\gg\delta$.
Assume the potential $V(\bx)$
  is radially symmetric and increasing in $r=|\bx|$, then the ground state $\phi_g\in S$ ($\phi_g^\vep$) can be chosen as a
radially symmetric function, which is decreasing in $r=|\bx|$. For $\beta\to-\infty$, we have
\be
\rho_g^\vep=|\phi_g^\vep|^2\to \rho_{\infty} \text{ in  } H^1,
\ee
where $\rho_\infty$ is the unique nonnegative minimizer of the energy \eqref{eq:eeta} in Theorem \ref{thm:fb}, which is radially symmetric and increasing in $r=|\bx|$.

(2) For case $2^\prime$, i.e. $\beta\to -\infty$ and 
 $\lim_{\beta\to-\infty}\delta/|\beta|^{\frac{4+d}{2+d}}=\delta_\infty>0$,
set 
$\vep=|\beta|^{-\frac{1}{2+d}}$. For $\beta\to-\infty$ ($\vep\to0^+$),
 there exists a subsequence $\beta_n\to-\infty$ 
such that for $\vep_n=|\beta_n|^{-\frac{1}{2+d}}\to0^+$ 
\be
\rho_g^{\vep_n}(\bx)\to \rho_g(\bx) \text{ in } H^1,
\ee
 where $\rho_g(\bx)$ is a nonnegative minimizer of the energy \eqref{eq:edelta}.

(3) For case $3^\prime$, i.e. $\beta\to-\infty$ and $\delta/|\beta|^{\frac{4+d}{2+d}}\gg1$, set 
$\vep=\delta^{-\frac{1}{4+d}}$. For $\delta\to+\infty$ ($\vep\to 0^+$), we have
\be
\rho_g^\vep(\bx)=|\phi_g^\vep(\bx)|^2\rightarrow\rho_\infty(\bx)=|\phi_\infty(\bx)|^2 \text{ in } H^1,
\ee
where $\phi_\infty(\bx)$
is the unique nonnegative minimizer of the energy \eqref{eq:es1}.
\end{theorem}
{\noindent\it Proof}.

(1)  Choosing $\vep=|\beta|^{1/2}/\delta^{1/2}$ in \eqref{eq:energyeps}, we find $\phi_g^\vep$ minimizes the following energy
\begin{equation}\label{eq:energyeta}
E_\eta(\phi)=
\int_{\Bbb R^d}\left[\frac{\eta_1}{2}|\nabla\phi|^2+\eta_2V(\bx)|\phi|^2-\frac12|\phi|^4+\frac12\left|\nabla|\phi|^2\right|^2\right]\,d\bx,\quad \phi\in S,
\end{equation}
with $\eta_1=\frac{\delta^{\frac{d-2}{2}}}{|\beta|^{d/2}}=o(1)$ and $\eta_2=\frac{\delta^{\frac{d+2}{2}}}{|\beta|^{\frac{4+d}{2}}}=o(1)$ when $\beta\to-\infty$
and $\delta=o(|\beta|^{\frac{4+d}{2+d}})$. Intuitively, only the leading $O(1)$ terms in \eqref{eq:energyeta} are important in the limit as $\beta\to-\infty$.
Under the hypothesis of a radially  increasing potential $V(\bx)$, we have (regularize $\phi_\infty=\sqrt{\rho_\infty}$ such that $\phi_\infty\in H^1$ if necessary)
\be
E_{1^\prime}(\sqrt{\rho_\infty})\le E_{1^\prime}(\sqrt{\rho_g^\vep})\leq E_\eta(\phi_g^\vep)\leq E_\eta(\sqrt{\rho_\infty})\leq o(1)+E_{1^\prime}(\sqrt{\rho_\infty}),
\ee
which shows $\lim\limits_{\beta\to-\infty}E_{1^\prime}(\sqrt{\rho_g^\vep})=E_{1^\prime}(\sqrt{\rho_\infty})=I_1$.
Similar to the proof of part (4) in Theorem \ref{thm:fb}, by using the radially decreasing property of $\rho_g^\vep$ which is a minimizing sequence for the energy $E_{1^\prime}$, it is straightforward to check that
as $\beta\to-\infty$,  $\rho_g^\vep$ converges to some radially decreasing function $\rho_0$ weakly in $H^1$ and strongly in $L^2$, $\|\nabla\rho^\vep_g\|\to\|\nabla\rho_0\|$ and $\|\rho_0\|_{L^1}=1$. Therefore, $\rho_g^\vep\to \rho_0$ in $H^1$ and $\rho_0$ is the unique radially decreasing minimizer of energy $E_{1^\prime}$ \eqref{eq:eeta}.

(2) Let $\vep=|\beta|^{-1/(d+2)}$ and $\rho_g^\vep(\bx)=\vep^{-d}|\phi_g(\bx/\vep)|^2$ where $\phi_g(\bx)$ is a ground state of \eqref{eq:minimize}, then
$\sqrt{\rho_g^{\vep}}\in S$ is a ground state of \eqref{eq:energyeps}. Using Nash inequality with the fact $\sqrt{\rho_g^{\vep}}\in S$, we can easily find
\be
\int_{\Bbb R^d}V(\bx)\rho_g^\vep(\bx)\,d\bx+\|\nabla\rho_g^\vep\|+\|\rho_g^\vep\|\leq C.
\ee
We can extract a subsequence $\vep_n\to0$, such that for some $\rho_0\in H^1$,
we have
\be
\rho_g^{\vep_n}\to \rho_0,\quad \text{weakly in } H^1,\quad \text{weakly}-\star\quad \text{in } L_V^1=\{\rho|\int_{\Bbb R^d}V(\bx)|\rho|\,dx<+\infty\},
\ee
and
\begin{equation*}
\int_{\Bbb R^d}V(\bx)\rho_0(\bx)+\|\nabla\rho_0\|+\|\rho_0\|\leq\liminf\limits_{\vep_n\to0^+}\left(\int_{\Bbb R^d}V(\bx)\rho_g^{\vep_n}(\bx)+\|\nabla\rho_g^{\vep_n}\|+\|\rho_g^{\vep_n}\|\right).
\end{equation*}
We then show that the convergence is strong in $L^2$. For any $\eta>0$, there exists $R>0$ such that $\int_{|\bx|>R}\rho_g^{\vep_n}(\bx)\,d\bx<\eta$ (confining property of $V(\bx)$).
Since $H^1(B_R)\hookrightarrow L^2(B_R)$ is compact, $\int_{|\bx|\leq R}|\rho_g^{\vep_n}(\bx)-\rho_0(\bx)|^2\,d\bx\to0$ and
\begin{align*}
&\limsup\limits_{\vep_n\to0}\|\rho_g^{\vep_n}-\rho_0\|^2\\
=&
\limsup\limits_{\vep_n\to0}\int_{|\bx|\leq R}|\rho_g^{\vep_n}(\bx)-\rho_0(\bx)|^2\,d\bx+\limsup\limits_{\vep_n\to0}\int_{|\bx|> R}|\rho_g^{\vep_n}(\bx)-\rho_0(\bx)|^2\,d\bx\\
\leq&\limsup\limits_{\vep_n\to0}\left(\int_{|\bx|> R}|\rho_g^{\vep_n}(\bx)-\rho_0(\bx)|\,d\bx\right)^{\frac12}
\left(\int_{|\bx|> R}|\rho_g^{\vep_n}(\bx)-\rho_0(\bx)|^3\,d\bx\right)^{\frac12}\\
\leq& C\eta^{1/2}.
\end{align*}
Hence $\limsup\limits_{\vep_n\to0}\|\rho_g^{\vep_n}-\rho_0\|^2=0$ and $\rho_g^{\vep_n}\to\rho_0$ in $L^2$, which implies that $\rho_0(\bx)\ge0$. Similarly, due to the confining property of
$V(\bx)$, $\|\rho_0\|_{L^1}=1$. In particular, regularizing the minimizer of $E_{2^\prime}(\cdot)$ \eqref{eq:edelta} if necessary, we have
\begin{equation*}
E_{2^\prime}(\sqrt{\rho_0})\leq \liminf\limits_{\vep^n\to0} E_{2^\prime}(\sqrt{\rho_g^{\vep_n}})\leq \limsup\limits_{\vep^n\to0} E^{\vep_n}(\sqrt{\rho_g^{\vep_n}})\leq
E^{\vep_n}(\sqrt{\rho_0}),
\end{equation*}
and $E^{\vep_n}(\sqrt{\rho_0})\leq E_{2^\prime}(\sqrt{\rho_0})+o(1)$, which verifies $\rho_0$ is a minimizer of $E_{2^\prime}(\cdot)$ \eqref{eq:edelta} as well as $\|\nabla\rho_g^{\vep_n}\|\to\|\nabla\rho_0\|$. Thus, $\rho_g^{\vep_n}\to\rho_0$ in $H^1$.

(3) The proof is similar to part (1), in view of the fact that the minimizer of \eqref{eq:es1} is unique.
$\hfill\Box$
%

\subsection{Vanishing higher order effect}
In this subsection, we consider the case $\delta\to0^+$, i.e. the vanishing higher order effects.
For fixed $\beta$, we denote $\phi_g^{\delta}$ to be the non-negative ground state corresponding to $\delta>0$.
When $\delta=0$, the MGPE degenerates to the GPE case, and the ground state exists \cite{Bao2013}
if $\beta\ge0$ when $d=3$, or $\beta>-C_b$ when $d=2$, or $\beta\in\Bbb R$ when $d=1$,   and $C_b$ is
defined as \cite{Weinstein,Guo}
\be\label{bestc}
C_b:=\inf_{0\ne f\in H^1({\mathbb R}^2)} \frac{\|\nabla
f\|_{L^2(\mathbb R^2)}^2\|f\|_{L^2(\mathbb R^2)}^2}{\|f\|_{L^4(\mathbb R^2)}^4}=\pi\cdot (1.86225\ldots).
\ee
It is obvious that, in such cases, there exists a subsequence $\delta_n\to0$ ($n=1,2,\ldots$), such that $\phi_g^{\delta_n}(\bx)\to\phi_g(\bx)$ in $H^1$,
where $\phi_g(\bx)$ is a  nonnegative minimizer of the energy
\be\label{eq:gpenergy}
E_{\rm GP}(\phi)=\int_{\Bbb R^d}\left[\frac{1}{2}|\nabla\phi|^2+V(\bx)|\phi|^2+\frac{\beta}{2}|\phi|^4\right]\,d\bx,\quad \|\phi\|=1.
\ee
Moreover, when $\beta\ge0$, the nonnegative minimizer $\phi_g$ of \eqref{eq:gpenergy} is unique and $\phi_g^\delta\to\phi_g$  in $H^1$ as $\delta\to0^+$.

A more interesting topic would be to study the cases for $\beta$ in the regimes where
the ground state does not exist when $\delta=0$.
In such cases, it is worth noticing that the ground state profiles will
have certain blow-up phenomenon  as $\delta\to0^+$, i.e., the density will concentrate towards a Dirac function.
This phenomenon can be characterized by the following theorem.
\begin{theorem}\label{thm:delta}($\delta\to0^+$ limit)
Let $V(\bx)$ ($\bx\in\Bbb R^d$, $d=1,2,3$) be given in \eqref{eq:potential:ho}, $\delta>0$, $\phi_g^\delta\in S$ be a nonnegative ground state of \eqref{eq:minimize}.

(1) When $d=2$ and $\beta<-C_b$, denoting $\tphi_{\delta}(\bx)=\sqrt{\delta}\phi_g^{\delta}(\sqrt{\delta}\mathbf{x})$, there exists a subsequence 
$\delta_n\to0$ such that
\be
\tphi_{\delta_n}(\bx)\to\phi_0(\bx) \text{ in } H^1,
\ee
where
 $\phi_0(\bx)$ is a  nonnegative minimizer of the energy
\be\label{eq:energybeta}
E_{\rm \beta}(\phi)=\int_{\Bbb R^d}\left[\frac{1}{2}|\nabla\phi|^2+\frac{\beta}{2}|\phi|^4+\frac{1}{2}|\nabla|\phi|^2|^2\right]\,d\bx, \text{ subject to  } \|\phi\|=1.
\ee

(2) When $d=3$, $\beta<0$ and $V(\bx)$ is radially  increasing, 
 the ground state  $\phi_g^\delta(\bx)$ can be chosen
 as  decreasing radially symmetric functions.  Let $\trho_{\delta}(\mathbf{x})=|\tphi_\delta(\bx)|^2$ , where $\tphi_{\delta}(\mathbf{x})=\delta^{3/4}\phi_g^{\delta}(\sqrt{\delta}\mathbf{x})$. For $\delta\to0^+$, we have
\be
\trho_{\delta}\to\rho_0 \text{ in } H^1,
\ee
where $\rho_0$ is the unique decreasing radially symmetric nonnegative minimizer
of the following energy
\be
E_r^{\beta}(\sqrt{\rho})=\int_{\Bbb R^d}\left[\frac{\beta}{2}|\rho|^2+\frac{1}{2}|\nabla\rho|^2\right]\,d\bx,\text{ with } \rho\ge0 \text{ and }  \int_{\Bbb R^d}\rho(\bx)\,d\bx=1.
\ee
More precisely, $\rho_{0}\ge0$ satisfies the free boundary problem
\be
\beta\rho-\Delta\rho=\mu\chi_{\rho>0}, \quad \rho|_{\partial\{\rho>0\}}=|\nabla\rho|\big|_{\partial\{\rho>0\}}=0,
\ee
where $\mu=2E_r^\beta(\sqrt{\rho_0})$.
\end{theorem}
{\noindent\it Proof}.

(1) The existence of the nonnegative minimizer of $E_\beta(\cdot)$ can be proved by a similar argument in Theorem \ref{thm:fb} for energy $E_{1^\prime}(\cdot)$ and the detail is omitted here. We denote the minimum energy of $E_\beta(\cdot)$ as $E_0$.

Letting $\vep=\delta^{-1/2}$ in \eqref{eq:energyeps}, it is obvious that $\tilde{\phi}_\delta(\bx)\in S$ minimizes the energy
\begin{equation}
\tilde{E}_{\delta}(\phi)=\int_{\Bbb R^d}\left[\frac{1}{2}|\nabla\phi|^2+\delta^2V(\bx)|\phi|^2+\frac{\beta}{2}|\phi|^4+\frac{1}{2}|\nabla|\phi|^2|^2\right]\,d\bx,\quad \phi\in S.
\end{equation}
Now, choosing a ground state $\phi_g\in S$ of \eqref{eq:energybeta} as a testing state (using a $C_0^\infty$ approximation if necessary for the potential term), we have
\begin{equation*}
\delta^2\int_{\Bbb R^d}V(\bx)|\tilde{\phi}_\delta(\bx)|^2\,d\bx+E_\beta(\tilde{\phi}_\delta)=\tilde{E}_{\delta}(\tilde{\phi}_\delta)\leq \tilde{E}_{\delta}(\phi_g)
\leq E_0+C\delta^2,
\end{equation*}
which implies $\int_{\Bbb R^d}V(\bx)|\tilde{\phi}_\delta(\bx)|^2\,d\bx\leq C$. Therefore, we have
\begin{equation*}
\int_{\Bbb R^d}V(\bx)|\tilde{\phi}_\delta(\bx)|^2\,d\bx+\|\tilde{\phi}_\delta\|_{H^1}+\|\nabla|\tilde{\phi}_\delta|^2\|\leq C.
\end{equation*}
Following the proof in Theorem \ref{thm:mres}, there exists $\phi_0\in H^1$ with $\|\phi_0\|=1$ and a subsequence $\delta_n\to0$ such that
$\phi_{\delta_n}\to\phi_0$ strongly in $L^2$ and weakly in $H^1$,
\begin{equation*}
E_{\beta}(\phi_0)\leq \liminf_{n\to\infty}E_{\beta}(\tilde{\phi}_{\delta_n})\leq\liminf_{n\to\infty}\tilde{E}_{\delta}(\tilde{\phi}_{\delta_n})
\leq E_0,
\end{equation*}
and $\phi_0$ is a minimizer of \eqref{eq:energybeta}. From the above inequality, it is easy to find that $\|\nabla\phi_{\delta_n}\|\to\|\nabla\phi_0\|$
 and thus $\phi_{\delta_n}\to\phi_0$ strongly in $H^1$.

(2) The proof is essentially presented in Theorem \ref{thm:tfneg}, part (1) and Theorem \ref{thm:fb}, part (4). $\hfill\Box$

\section{Limiting behavior of ground states in  bounded domains}\label{subsec:bd}

Now we consider \eqref{eq:mgpe} defined in a bounded domain $\Omega\subset\Bbb R^d$,
the limiting profiles of ground states \eqref{eq:minimize} are considered under different sets of parameters $\delta$
and $\beta$. To simplify the discussion, we choose
the external potential as box potential, i.e.
\be\label{eq:boxpt}
V(\bx)=\begin{cases}0,&\bx\in\Omega,\\
+\infty,&\text{otherwise}.\end{cases}
\ee
The energy $E(\cdot)$ \eqref{eq:energy} reduces to
\be\label{eq:energybd}
E_{\Omega}(\phi)=\int_{\Omega}\biggl[
\frac12|\nabla\phi|^2+\frac{\beta}{2}|\phi|^4+\frac{\delta}{2}\left|\nabla|\phi|^2\right|^2\biggl]\,d\bx,
\ee
and the ground state $\phi_g$ is then the minimizer of the energy $E_{\Omega}$ under the constraint $\|\phi\|_{L^2(\Omega)}=1$.
 The major difference between
the whole space case (section \ref{subsec:ws}) and the bounded domain case is that the scalings under different sets of parameters are very different. We list the following different regimes for bounded
domain:

 Case B1: $\beta\to +\infty$ and $\delta=o(\beta)$;\\
 \hspace*{.4cm} Case B2: $\beta\to+\infty$ and $\lim_{\beta\to+\infty}\delta/\beta=\delta_\infty>0$;\\
 \hspace*{.4cm} Case B3: $\beta\to+\infty$ and  $\delta/\beta\gg1$, i.e. $\beta=o(\delta)$ as $\delta\to+\infty$;

\noindent and for $\beta\to-\infty$

 Case B$1^\prime$: $\beta\to -\infty$ and  $\delta=o(\beta)$;\\
 \hspace*{.4cm} Case B$2^\prime$: $\beta\to-\infty$ and $\lim_{\beta\to-\infty}\delta/|\beta|=\delta_\infty>0$;\\
 \hspace*{.4cm} Case B$3^\prime$: $\beta\to-\infty$ and  $\delta/|\beta|\gg1$, i.e. $|\beta|=o(\delta)$ as $\delta\to+\infty$.

\begin{theorem}\label{thm:tfbd}
(Thomas-Fermi limit) Let $V(\bx)$ ($\bx\in \Omega$, $d=1,2,3$) be a box potential, $\delta>0$, and $\phi_g\in S$ be the positive ground state of \eqref{eq:minimize}.

(1) For Case B1, i.e. $\beta\to +\infty$ and $\delta=o(\beta)$,
we have
\be
\rho_g^{\beta}=|\phi_g(\bx)|^2\to\rho_\infty(\bx):=|\phi_\infty(\bx)|^2 \text{ in } L^2,
\ee
where $\phi_\infty(\bx)$ is the unique nonnegative minimizer of the energy
\be\label{}
E_b(\phi)=\int_{\Omega}\frac{1}{2}|\phi|^4\,d\bx\text{ with }  \|\phi\|^2=1.
\ee
More precisely, $\rho_\infty=\frac{1}{|\Omega|}$ 
with $|\Omega|$ being the volume of the domain.

(2) For Case B2, i.e. $\beta\to +\infty$ and $\lim\limits_{\beta\to+\infty}\delta/\beta=\delta_\infty>0$ for some $\delta_\infty>0$,
we have
\be
\rho_g^{\beta,\delta}=|\phi_g(\bx)|^2\to\rho_\infty(\bx) \text{ in } H^1,
\ee
where $\rho_\infty(\bx)$ is the unique nonnegative minimizer of the energy
\be
E_{bd}^+(\sqrt{\rho})=\int_{\Omega}\left[\frac{1}{2}|\rho|^2+\frac{\delta_\infty}{2}|\nabla\rho|^2\right]\,d\bx,\text{ where } \rho\ge0 \text{ and } \int_{\Omega}\rho(\bx)\,d\bx=1.
\ee
Moreover, $\rho_{\infty}\ge0$ satisfies the equation
\be
\rho-\delta_\infty\Delta\rho=2E_{bd}^+(\sqrt{\rho}) \text{ for } \bx\in\Omega, \text{ and }\rho(\mathbf{x})|_{\partial\Omega}=0.
\ee

(3) For Case B3, i.e. $\delta\to +\infty$ and $\beta=o(\delta)$, we have
\be
\rho_g^{\delta}=|\phi_g(\bx)|^2\to\rho_\infty(\bx) \text{ in } H^1,
\ee
where $\rho_\infty(\bx)$ is the unique nonnegative minimizer of the energy
\be
E_d(\sqrt{\rho})=\int_{\Omega}\frac{1}{2}|\nabla\rho|^2\,d\bx,\text{ where } \rho\ge0\text{ and }\int_{\Omega}\rho(\bx)\,d\bx=1.
\ee
Moreover, $\rho_{\infty}\ge0$ satisfies the equation
\be
-\Delta\rho=2E_d(\sqrt{\rho}) \text{ for } \bx\in\Omega, \text{ and }\rho(\mathbf{x})|_{\partial\Omega}=0.
\ee
\end{theorem}
{\noindent\it Proof}.
The proofs are similar to those in Theorem \ref{thm:TFp}. $\hfill\Box$
\begin{remark}
\begin{itemize}
 \item In Theorem \ref{thm:tfbd}, part (3)  holds true for Case B$3^\prime$, i.e. $\beta\to-\infty$ and $\delta\gg|\beta|$.
 \item For Case B$2^\prime$,
i.e. $\beta\to-\infty$ and  $\lim\limits_{\beta\to-\infty}\delta/|\beta|=\delta_\infty>0$ for some $\delta_\infty>0$,
we have a subsequence $\beta_n\to-\infty$ and $\delta_n$ such that
\be
\rho_g^{\beta_n,\delta_n}=|\phi_g^{\beta_n,\delta_n}(\bx)|^2\to\rho_\infty(\bx) \text{ in } H^1,
\ee
where $\rho_\infty(\bx)$ is a  nonnegative minimizer of the energy
\be
E_{bd}^-(\sqrt{\rho})=\int_{\Omega}\left[-\frac{1}{2}|\rho|^2+\frac{\delta_\infty}{2}|\nabla\rho|^2\right]\,d\bx,\text{ where } \rho\ge0\text{ and }\int_{\Omega}\rho(\bx)\,d\bx=1.
\ee
More precisely, $\rho_{\infty}\ge0$ satisfies the equation
\be
\rho_\infty-\delta_0\Delta\rho_\infty=2E_{bd}^-(\sqrt{\rho_\infty})\chi_{\{\rho_\infty>0\}} \text{ for } \bx\in\Omega, \text{ and }\rho_\infty(\mathbf{x})|_{\partial\Omega}=0.
\ee
\end{itemize}
\end{remark}
It remains to consider the last case B$1^\prime$   as $\beta\to-\infty$ and $\delta=o(|\beta|)$. For simplicity, we
assume $\Omega$ is a ball in $\Bbb R^d$.

\begin{theorem}
Consider the box potential given in \eqref{eq:boxpt} with $\Omega=B_R:=\{|\bx|<R\}$. For case B$1^\prime$, i.e. $\beta\to-\infty$, $\delta>0$ and $\delta=o(|\beta|)$, 
the ground state of \eqref{eq:energybd}, denoted as $\phi_g^{\beta,\delta}\in H^1(\Bbb R^d)$, can be chosen to be a non-increasing radially symmetric function.
Let $\phi_g^\vep(\bx)=\vep^{d/2}\phi_g^{\beta,\delta}(\bx\vep)\in S$
with $\varepsilon=\delta^{1/2}/|\beta|^{1/2}$. Then $\vep\to0^+$ as $\beta\to-\infty$,
and we have
\be
\rho_g^\vep=|\phi_g^\vep|^2\to \rho_{\infty} \text{ in } H^1,
\ee
where $\rho_\infty$ is the unique non-increasing radially symmetric minimizer of energy $E_{1^\prime}$ \eqref{eq:eeta}.
\end{theorem}
{\noindent\it Proof}.
 Let $\Omega^\vep=\{\bx/\vep,\,\bx\in\Omega\}$.
Since $\rho_\infty$ is compactly supported as shown in Theorem \ref{thm:fb}, for sufficiently small $\vep$, we have $\text{supp}(\rho_\infty)\subset \Omega^\vep$.
On the other hand,  $\phi_g^\vep$ minimizes the energy
\be
E_{\rm box}^\eta(\phi)=\int_{\Bbb R^d}\biggl[
\frac{\eta}{2}|\nabla\phi|^2-\frac{1}{2}|\phi|^4+\frac{1}{2}\left|\nabla|\phi|^2\right|^2\biggl]\,d\bx,\quad \phi\in H_0^1(\Omega^\vep),\quad \|\phi\|=1,
\ee
where $\eta=\frac{\delta^{\frac{d-2}{2}}}{|\beta|^{\frac{d}{2}}}=o(1)$ as $\vep\to0^+$.  We can then proceed as that in Theorem \ref{thm:tfneg} and
the limit of $\rho_g^\vep$ as $\beta\to-\infty$ ($\vep\to0^+$) follows. $\hfill\Box$

Similarly, we could extend the $\delta\to0^+$ limit results in Theorem \ref{thm:delta} to the bounded domain case here. Since no different scaling  is involved, the extension is
straightforward and we omit it here.

\section{Conclusion}
We  have analyzed the ground state of a Bose-Einstein condensate in the presence of higher-order interaction (HOI),
modeled by a modified Gross-Pitaevskii equation (MGPE).
The ground state structures are quite different from the case without HOI.
We established the existence and  uniqueness as well as non-existence
results on ground states in different parameter regimes. The asymptotic profiles of the ground states
under different combinations of the HOI and the contact interaction were studied. The limiting profiles were found to
be quite interesting and complicated involving free boundary problems. 

\section*{Acknowledgments}
This work was partially supported by the Academic Research
Fund of Ministry of Education of Singapore grant No.
R-146-000-223-112 (W. Bao and X. Ruan), the NSFC grant No. U1530401 (Y. Cai), 11771036 (Y. Cai)  and  91630204 (Y. Cai). This work was partially done
when the authors were visiting the Institute for Mathematical Sciences
at the National University of Singapore in 2018.




\section*{\refname}
\begin{thebibliography}{}
\bibitem{Anderson}
{\sc M. H. Anderson, J. R. Ensher, M. R. Matthewa, C. E.
Wieman and E. A. Cornell},
\textit{Observation of Bose-Einstein
condensation in a dilute atomic vapor}, Science, 269 (1995),
pp.~198--201.

\bibitem{Antoine}
{\sc X. Antoine, W. Bao and C. Besse},
\textit{Computational methods for the dynamics of the nonlinear Schr\"odinger/Gross-Pitaevskii equations},
Comput. Phys. Commun.,  184 (2013), pp.~2621-2633.

\bibitem{Aschbacher}
{\sc W. Aschbacher, J. Fr\"{o}hlich, G. Graf, K. Schnee, and M. Troyer},
 \textit{Symmetry breaking regime in the nonlinear Hartree equation},
 J. Math. Phys., 43 (2002), pp.~3879--3891.

\bibitem{Bao2014}
{\sc W. Bao},
\textit{Mathematical models and numerical methods for Bose-Einstein condensation},
Proceeding of International Congress of Mathematicians (Seoul 2014), IV (2014), pp. 971-996.


\bibitem{Bao1}
{\sc W. Bao}, \textit{Ground states and dynamics of multicomponent
Bose--Einstein condensates}, Multiscale Model. Simul., 2 (2004),
pp.~210--236.

\bibitem{Bao20009}
{\sc W. Bao and Y. Cai},
\textit{Ground states of two-component Bose-Einstein condensates with an internal atomic Josephson junction},
East Asia J. Appl. Math., 1 (2010), pp.~49--81.

\bibitem{Bao2013}
{\sc W. Bao and Y. Cai},
\textit{Mathematical theory and numerical methods for Bose-Einstein condensation},
Kinet. Relat. Mod.,  6 (2013), pp.~1--135.

\bibitem{Wz1}
{\sc W. Bao and Q. Du}, \textit{Computing the ground state solution
of Bose-Einstein condensates by a normalized gradient flow}, SIAM J.
Sci. Comput., 25 (2004), pp.~1674-1697.

\bibitem{BaoDuZhang}
{\sc W. Bao, Q. Du and Y. Zhang}, \textit{Dynamics of rotating Bose-Einstein condensates and its efficient and accurate numerical computation}, SIAM J. Appl. Math., 66 (2006), pp.~758-786.



\bibitem{Bao2003}
{\sc W. Bao, D. Jaksch and P. A. Markowich},
\textit{Numerical solution of the Gross-Pitaevskii equation for Bose-Einstein condensation},
J. Comput. Phys., 187 (2003), pp.~318 - 342.





\bibitem{Ben}
{\sc N. Ben Abdallah, F. M\'ehats, C. Schmeiser and R. M. Weish\"aupl}
\textit{The nonlinear Schr\"odinger equation with a strongly anisotropic harmonic potential},
SIAM J. Math. Anal., 47 (2005), pp. 189--199.

\bibitem{Borov}
{\sc A. V. Borovskii and A. L. Galkin},
\textit{Dynamical modulation of an ultrashort high-intensity laser
pulse in matter}, JETP, 77 (1993), pp. 562--573.
\bibitem{Collin}
{\sc A. Collin, P. Massignan and C. J. Pethick},
\textit{Energy-dependent effective interactions for dilute many-body systems},
Phys. Rev. A,  75 (2007), 013615.
\bibitem{ColinM}
{\sc M. Colin, L. Jeanjean and  M. Squassina},
\textit{Stability and instability results for standing waves of quasi-linear Schr\"odinger equations},
Nonlinearity,  23 (2010), pp.~1353--1385.


\bibitem{Davis}
{\sc K. B. Davis, M. O. Mewes, M. R. Andrews, N. J. van Druten,
 D. S. Durfee, D. M. Kurn and W. Ketterle},
\textit{Bose-Einstein condensation in a gas of sodium atoms}, Phys.
Rev. Lett., 75 (1995), pp. 3969--3973.
\bibitem{De}
{\sc A. De Bouard, N. Hayashi,  and J.C. Saut},
\textit{Global existence of small solutions to a relativistic
nonlinear Schr\"odinger equation},
Comm. Math. Phys., 189 (1997), pp.~73-105.
\bibitem{Esry}
{\sc B. D. Esry and C. H. Greene},
\textit{Validity of the shape-independent approximation for Bose-Einstein condensates},
Phys. Rev. A,  60 (1999), 1451--1462.
\bibitem{Evans}
{\sc L. C. Evans},
\textit{Partial Differential Equations},  Amer.
Math. Soc., Providence, RI, 1998.
\bibitem{Fu}
{\sc H. Fu, Y. Wang and B. Gao},
\textit{Beyond the Fermi pseudopotential: A modified Gross-Pitaevskii equation},
Phys. Rev. A,  67 (2002), 053612.

\bibitem{Guo1}
{\sc Y. Guo and R. Seiringer},
\textit{Symmetry breaking and collapse in Bose-Einstein condensates with attractive interactions},
Lett. Math. Phys., 104 (2014), pp.~141--156.

\bibitem{Guo}
{\sc Y. Guo, X. Zeng and H. Zhou},
\textit{Energy estimates and symmetry breaking in attractive Bose-Einstein condensates with ring-shaped potentials},
Annales de l'Institut Henri Poincare (C) Non Linear Analysis, 33 (2016),  809--828,
\bibitem{Han}
{\sc Q. Han and F. H. Lin},
\textit{Elliptic Partial Differential Equations}, Amer. Math.
Soc., 2nd ed., 2011.

\bibitem{Kuri}
{\sc S. Kurihura},
\textit{Large-amplitude quasi-solitons in superfluid films},
J. Phys. Soc. Jpn, 50 (1981), pp.~3262--3267.




\bibitem{LiebL}
{\sc E. H. Lieb and M. Loss},
\textit{Analysis, Graduate Studies in Mathematics}, Amer. Math.
Soc., 2nd ed., 2001.
\bibitem{Lie}
{\sc E. H. Lieb, R. Seiringer and J. Yngvason}, \textit{Bosons in a
trap: a rigorous derivation of the Gross-Pitaevskii energy
functional}, Phy. Rev. A,  61 (2000), article 043602.
\bibitem{Liu}
{\sc J.Q. Liu, Y.Q. Wang, Z.Q. Wang},
\textit{Soliton solutions for quasi-linear Schr\"odinger equations
II},
J. Differential Equations, 187 (2003), pp.~473--493.

\bibitem{Liu2}
{\sc J.Q. Liu, Y.Q. Wang, Z.Q. Wang},
\textit{Solutions for quasilinear Schr\"odinger equations via the Nehari method},
Comm. Partial Differential Equations, 29 (2004), pp.~879--901.

\bibitem{LuG}
{\sc G. Lu  and B. Ou},
\textit{A Poincar{\'e} inequality on $R^n$ and its application to potential fluid flows in space},
Comm. Appl. Nonlinear Anal, 12 (2005), pp.~1--24.

\bibitem{Lu}
{\sc J. Lu and J. L. Marzuola},
\textit{Strang splitting methods for a quasilinear Schr\"odinger equation - convergence, instability and dynamics}, Commun. Math. Sci.,
13(2015), pp. 1051-1074.
\bibitem{Maris}
{\sc M. Maris},
\textit{On the symmetry of minimizers},
Arch. Ration. Mech. Anal.,  192 (2009), pp.~311--330.

\bibitem{Marz}
{\sc J.Marzuola, J.Metcalfe and D. Tataru},
\textit{Quasilinear Schr\"odinger equations II: Small data and cubic nonlinearities},
 Kyoto J. Math., 54 (2014), pp.~529--546.

\bibitem{Petro}
{\sc A. Petrosyan, H. Shahgholian, and N. N. Uraltseva},
 \textit{Regularity of Free
Boundaries in Obstacle-Type Problems}, AMS, 2012.
\bibitem{PitaevskiiStringari}
{\sc L. P. Pitaevskii and S. Stringari},
\textit{Bose-Einstein Condensation},
Clarendon Press, Oxford, 2003.
\bibitem{Popp}
{\sc M. Poppenberg},
\textit{On the local well posedness of quasi-linear Schr\"odinger equations in arbitrary
space dimension},
J. Differential Equations, 172 (2001), pp.~83--115.

\bibitem{Qi}
{\sc W. Qi, Z. Liang and Z. Zhang},
\textit{The stability condition and collective
excitation of a trapped Bose-Einstein
condensate with higher-order interactions},
J. Phys. B: At. Mol. Opt. Phys., 46 (2013), 175301.

\bibitem{Qix}
{\sc X. Qi and X. Zhang},
\textit{Modulational instability of a modified Gross-Pitaevskii equation with higher-order nonlinearity},
Phys. Rev. E, 86 (2012), 017601.

\bibitem{Ruan}
{\sc X. Ruan, Y. Cai and W. Bao},
\textit{Mean-field regime and Thomas-Fermi approximations of trapped Bose-Einstein condensates with higher order interactions in one and two dimensions},
J. Phys. B: At. Mol. Opt. Phys., 49 (2016), 125304.

\bibitem{Takeno}
{\sc S. Takeno and S. Homma},
\textit{Classical planar Heisenberg ferromagnet, complex scalar fields and nonlinear
excitations}, Progr. Theoret. Phys., 65 (1981), pp.~172--189.

\bibitem{Tho}
{\sc M. Th{\o}gersen, N. T. Zinner and A. S. Jensen},
\textit{Thomas-Fermi approximation for a condensate with higher-order interactions},
Phys. Rev. A , 80 (2009), 043625.


\bibitem{Veksler}
{\sc H. Veksler, S. Fishman and W. Ketterle},
\textit{A simple model for interactions and corrections to the Gross-Pitaevskii Equation},
Phys. Rev. A, 90 (2014), 023620.


\bibitem{Weinstein}
{\sc M. I. Weinstein}, \textit{Nonlinear Schr\"{o}dinger equations and sharp interpolation estimates}, Comm.
Math. Phys., 87 (1983), pp.~567--576.



\end{thebibliography}
\end{document}